\documentclass[iop,numberedappendix,appendixfloats,twocolappendix,revtex4]{emulateapj}
\usepackage{amsmath, latexsym, color, verbatim}
\usepackage{graphicx,epsfig, natbib, epstopdf}
\usepackage{pdflscape, rotating}
\usepackage{animate}

\def\alwaysmath#1{\ifmmode{#1}\else{$#1$}\fi}

\lefthead{Law et al. 2021}
\righthead{Dynamics of Ionized Gas}

\slugcomment{DRAFT: \today}

\begin{document}

\newcommand{\msun}{\ensuremath{\rm M_\odot}}
\newcommand{\msunyr}{\ensuremath{\rm M_{\odot}\;{\rm yr}^{-1}}}
\newcommand{\Ha}{\ensuremath{\rm H\alpha}}
\newcommand{\Pa}{\ensuremath{\rm Pa\alpha}}
\newcommand{\Hb}{\ensuremath{\rm H\beta}}
\newcommand{\lya}{\ensuremath{\rm Ly\alpha}}
\newcommand{\Ntwo}{[\ion{N}{2}]}
\newcommand{\kms}{\textrm{km~s}\ensuremath{^{-1}\,}}
\newcommand{\ztwo}{\ensuremath{z\sim2}}
\newcommand{\zthree}{\ensuremath{z\sim3}}
\newcommand{\feh}{\textrm{[Fe/H]}}
\newcommand{\afeh}{\textrm{[$\alpha$/Fe]}}
\newcommand{\nifeh}{\textrm{[Ni/Fe]}}

\newcommand{\oone}{\textrm{[O\,{\sc i}]}}
\newcommand{\ofour}{\textrm{[O\,{\sc iiii}]}}
\newcommand{\othree}{\textrm{[O\,{\sc iii}]}}
\newcommand{\otwo}{\textrm{[O\,{\sc ii}]}}
\newcommand{\ntwo}{\textrm{[N\,{\sc ii}]}}
\newcommand{\stwo}{\textrm{[S\,{\sc ii}]}}

\newcommand{\redtxt}[1]{\textcolor{red}{#1}}

\title{SDSS-IV MaNGA: Refining Strong Line Diagnostic Classifications Using Spatially Resolved Gas Dynamics}

%The Relation between Kinematics and Strong-Line Ionization Diagnostics

\author{David R.~Law\altaffilmark{1}, 
Xihan Ji\altaffilmark{2},
Francesco Belfiore\altaffilmark{3}, 
Matthew A.~Bershady\altaffilmark{4,5,6}, 
Michele Cappellari\altaffilmark{7},
Kyle B.~Westfall\altaffilmark{8},
Renbin Yan\altaffilmark{2},
Dmitry Bizyaev\altaffilmark{9,10}, 
Joel R.~Brownstein\altaffilmark{11},
Niv Drory\altaffilmark{12},
Brett H.~Andrews\altaffilmark{13}
}

\altaffiltext{1}{Space Telescope Science Institute, 3700 San Martin Drive, Baltimore, MD 21218, USA; dlaw@stsci.edu}
\altaffiltext{2}{Department of Physics and Astronomy, University of Kentucky, 505 Rose Street, Lexington, KY 40506-0057, USA.}
\altaffiltext{3}{INAF -- Osservatorio Astrofisico di Arcetri, Largo E. Fermi 5, I-50157, Firenze, Italy.}
\altaffiltext{4}{University of Wisconsin - Madison, Department of Astronomy, 475 N. Charter Street, Madison, WI 53706-1582, USA.}
\altaffiltext{5}{South African Astronomical Observatory, PO Box 9, Observatory 7935, Cape Town, South Africa.}
\altaffiltext{6}{Department of Astronomy, University of Cape Town, Private Bag X3, Rondebosch 7701, South Africa.}
\altaffiltext{7}{Sub-department of Astrophysics, Department of Physics, University of Oxford, Denys Wilkinson Building, Keble Road, Oxford OX1 3RH, UK.}
\altaffiltext{8}{University of California Observatories, University of California, Santa Cruz, 1156 High St., Santa Cruz, CA 95064, USA.}
\altaffiltext{9}{Apache Point Observatory and New Mexico State
University, P.O. Box 59, Sunspot, NM, 88349-0059, USA}
\altaffiltext{10}{Sternberg Astronomical Institute, Moscow State
University, Moscow, Russia}
\altaffiltext{11}{University of Utah, Department of Physics and Astronomy, 115 S. 1400 E., Salt Lake City, UT 84112, USA}
\altaffiltext{12}{McDonald Observatory, The University of Texas at Austin, 2515 Speedway, Stop C1402, Austin, TX 78712, USA.}
\altaffiltext{13}{PITT PACC, Department of Physics and Astronomy, University of Pittsburgh, Pittsburgh, PA 15260, USA}

\begin{abstract}

We use the statistical power of the MaNGA integral-field spectroscopic galaxy survey to improve the definition of strong line diagnostic boundaries used to classify gas ionization properties in galaxies. We detect line emission from 3.6 million spaxels distributed across 7400 individual galaxies spanning a wide range of stellar masses, star formation rates, and morphological types, and find that the gas-phase velocity dispersion $\sigma_{\Ha}$ correlates strongly with traditional optical emission line ratios such as \stwo/\Ha, \ntwo/\Ha, \oone/\Ha, and \othree/\Hb. Spaxels whose line ratios are most consistent with ionization by galactic HII regions exhibit a narrow range of dynamically cold line of sight velocity distributions (LOSVDs) peaked around 25 \kms\ corresponding to a galactic thin disk, while those consistent with ionization by  active galactic nuclei (AGN) and low-ionization emission-line regions (LI(N)ERs) have significantly broader LOSVDs extending to 200 km s$^{-1}$.  Star-forming, AGN, and LI(N)ER regions are additionally well separated from each other in terms of their stellar velocity dispersion, stellar population age, \Ha\ equivalent width, and typical radius within a given galaxy. We use our observations to revise the traditional emission line diagnostic classifications so that they reliably identify distinct dynamical samples both in two-dimensional representations of the diagnostic line ratio space and in a multi-dimensional space that accounts for the complex folding of the star forming model surface. By comparing the MaNGA observations to the SDSS single-fiber galaxy sample we note that the latter is systematically biased against young, low metallicity star-forming regions that lie  outside of the 3 arcsec fiber footprint.

\end{abstract}

\keywords{galaxies: kinematics and dynamics --- galaxies: spiral --- galaxies: statistics --- techniques: imaging spectroscopy}

%%%%%%%%%%%%%%%%%%%%%%%%%%%%%%%%%%%%%%%%%%%

\section{Introduction}
\label{intro.sec}

Gas is a key component of galaxies, providing the reservoirs of fuel used to form
stars and produce the luminous stellar structures that we observe.  This gas is
present in a variety of forms and phases from the cold H$_2$ molecular gas in which
stars form to neutral HI gas reservoirs, to warm ionized gas in HII regions
and hot ionized gas in the circumgalactic medium.  Over the lifetime of a galaxy
gas continually cycles between these phases, collapsing into new stars and being
expelled by the end products of stellar evolution to begin the process again
\citep[see, e.g.,][and references therein]{tumlinson17, ph20}.

In the warm ionized gas,
emissive cooling is achieved via discrete emission from key atomic
transitions.  Following recombinations in the ionized hydrogen gas,
a part of this cooling comes from a cascade of energy level transitions through
key emission lines such as \Ha\ $\lambda 6564$, \Hb\ $\lambda 4863$, and others of the Lyman, Balmer, and Paschen series.
While less abundant, metals such as oxygen, nitrogen, and sulfur provide
another important avenue for cooling via forbidden line emission triggered
by collisional excitations within regions of neutral (e.g., \oone), singly ionized (e.g., \stwo), or 
multiply-ionized gas (e.g., \othree)
overlapping to various degrees with the ionized hydrogen.
The observed strength of a given line is a product of a variety of factors 
including the spectral shape and intensity of the ionizing radiation, the ionization
potential of a given ion (e.g., 10.4 eV for singly ionized S$^+$, 35.2 eV for doubly
ionized $O^{++}$), and the temperature, density, and metal abundance of the gas.
The relative strengths of different lines
thus provide sensitive diagnostics of the properties of the ionized
gas and the astrophysical sources illuminating them.

\citet[][hereafter `BPT']{baldwin81} popularized the analysis of
intensity ratios between optical emission lines that are close together in 
wavelength, and therefore both free from differential dust extinction
and easy to observe simultaneously.  \citet{vo87} and \citet{ho97}
further developed this framework, leading to some of the most widely used
diagnostic ratios including log(\ntwo\ $\lambda 6585$/\Ha)  (hereafter
`N2'), log(\stwo\ $\lambda 6718+6732$/\Ha) (hereafter 
`S2'),  
log(\oone\ $\lambda 6302$/\Ha) (hereafter 
`O1'), and log(\othree\ $\lambda 5008$/\Hb) (hereafter `R3').
\citet{dopita00} and \citet{kewley01} put such observational diagnostics
on a firm theoretical footing, using stellar photonionization models
to identify a series of relations \citep[see, e.g.,][and references therein]{kewley19}
that divide parameter space into 
regions consistent with `classical' HII regions resulting from young massive stars
and regions fueled by radiative shocks or emission from active galactic nuclei (AGN).

% Differences between the different diagnostics
Each of these emission-line diagnostics has its own strengths and weaknesses
\citep[see review by][and references therein]{maiolino19}.
For instance, the N2 line ratio correlates strongly with gas-phase metallicity in galactic HII regions; within the N2-R3 star-forming sequence the location of a given galaxy is governed largely by the metallicity (increasing
which moves a galaxy along the sequence to
smaller R3 at larger N2) while increasing the ionization parameter (i.e., the dimensionless ratio
between the number of hydrogen ionizing photons and the hydrogen density)
moves a galaxy across the sequence to larger R3 at larger N2. However, the
N2 diagnostic can also be biased by unusual variations in the nitrogen
abundance, and
does less well at distinguishing between other common sources of
ionizing photons.  In contrast, the S2 and O1 diagnostics are robust
against nitrogen abundance variation and more cleanly distinguish
between ionization sources ascribed to AGN and LI(N)ERs \citep[see, e.g.][]{dopita95}.
While the former category is relatively well understood to result from
energetic feedback derived from a central active nucleus, the LI(N)ER category is more of a historical classification corresponding to galaxies with strong low-ionization emission lines (e.g., \ntwo, \stwo, and \oone) 
and (compared to AGN) relatively weak higher ionization lines
such as \othree.
Although such regions were originally observed in the centers of galaxies \citep[e.g.,][]{heckman80},
recent observations have shown that LI(N)ER-type line ratios occur at larger galactocentric distances as well \citep{sarzi10} and display relatively flat ionization parameter gradients \citep{yan12}, suggesting that they are generally unrelated to AGN activity. \citet{belfiore16} further distinguished between `cLIER' galaxies (i.e., those with LI(N)ER-like emission at small radii and HII regions at larger radii) and `eLIER' galaxies (i.e., those with LI(N)ER-like emission throughout the galaxy and no evidence for significant star formation at any radius).

Complicating these categories is the presence of a substantial amount of \Ha\ emission (30\% or more of the integrated emission for a given galaxy) from so-called diffuse ionized gas (DIG), also known as the warm ionized medium (WIM), which in the Milky Way is known as the Reynolds layer \citep{reynolds90}.
Similar to LI(N)ERs, DIG emission exhibits enhanced N2, S2, and O1 line ratios with respect to galactic HII regions. Although the source of the ionizing photons illuminating the DIG is uncertain, \citet{zhang17} argue in favor of post-AGB stars as the dominant mechanism in general since radiation escaping from HII regions is not hard enough to produce the observed line ratios.

% Bridge statement: flux ratios are limited, and rely on theoretical models of the SOURCES of the ionizing
% photons.
% Dynamics are observable, and can tell us about the properties of the gas ILLUMINATED by these photons.
% We can use the latter to help constrain the former.

Consequently, while strong line ratios are easy to measure they can be challenging to interpret in terms
of the underlying physical origins of the ionizing photons
\citep[see, e.g., review by][]{sanchez20a}.  First and foremost, line ratios can vary dramatically within individual galaxies with gas in different regions ionized by a variety of different
mechanisms. Such spatial variations are lost in spectroscopic surveys that do not spatially resolve
individual regions within galaxies \citep[e.g., the original SDSS single-fiber survey,][]{k03,bman04,tremonti04}, resulting in line ratios produced by the blending of various components. Additionally, while stellar photoionization models exhibit a characteristic
folding of the ionization surface (see discussion in \S \ref{cloudy.sec})
the precise upper envelope of the star-forming sequence is confused by contributions from DIG and the overlap with
AGN and LI(N)ER models, further complicating efforts to separate different
physical mechanisms.

A key additional diagnostic can therefore be provided by spatially resolved integral field unit (IFU) spectroscopic measurements of the gas-phase velocity dispersion $\sigma_{\Ha}$. Star formation generally happens in HII regions with intrinsic thermal broadening $\sim 9$ \kms\ \citep[e.g.,][]{o89} and a potentially comparable
non-thermal expansion component; on kpc scales encompassing many such
regions the additional broadening provided by the velocity dispersion between individual HII
regions in the galactic thin disk produces an observed $\sigma_{\Ha} \sim 15-30$ \kms \citep[e.g.,][]{jv99,rozas00, rozas02,relano05,andersen06}. Other ionization mechanisms in contrast trace gas clouds with large-scale streaming motions
or extended three-dimensional distributions, and consequently possessing
different velocities that produce a 
broader integrated $\sigma_{\Ha}$ of 50 \kms\ or more.
We therefore expect the line of sight velocity distribution (LOSVD) to  directly measure the dynamical
properties of the ionized gas and thereby help to constrain the physical mechanisms likely to be ionizing that gas.

%francbelf: actually there is an interesting correlation between luminosity and velocity dispersion in giant HII regions (which some people have used to get constrains on cosmology http://www.ifa.hawaii.edu/users/kud/teaching_15/13_HII.pdf)
%francbelf: this is true in normal regions, but less true in regions of very high SFR/EWHA, where one may actually be dominated by one/a few giant HII regions

Multiple studies in the last decade have explored exactly such correlations.  Some of the earliest used
IFU observations of luminous and ultraluminous infrared galaxies ((U)LIRGs), with \citet{mi06, mi10} and \citet{maca06} noting the positive correlation between $\sigma_{\Ha}$ and the
N2, S2, and O1 line ratios, which they interpret as evidence for the contribution of shocks.
Similarly, \citet{rich11} found a bimodal distribution of velocity dispersions in a study of two nearby LIRGs with peaks above and below 50 \kms\ that they identified as corresponding to ionized gas in 
unresolved HII regions and high-velocity shocks respectively. More recently, \citet{oparin18}, \citet{da19}, and \citet{lopez20}
used the correlation between $\sigma_{\Ha}$ and the degree of excitation
observed in a handful of nearby galaxies to propose that $\sigma_{\Ha}$ be used as an additional diagnostic to help distinguish gas photoionized by HII regions from AGN and shock-driven excitation mechanisms \citep[see also][]{zhang18}.
Targeted studies using the MUSE instrument on the VLT have been able to push this work even further.
\citet{db20} for instance noted that the DIG in NGC 7793 had a higher amount of turbulence than the HII regions, while \citet{brok20} used observations of 41 star-forming galaxies to note that the asymmetric drift of the DIG suggested that it was distributed throughout a  layer thicker than the star-forming disk but thinner than the stellar disk.

In order to characterize such correlations in detail however, it is necessary to employ large and representative
galaxy samples that cover both the star-forming and quiescent galaxy sequences.
Some of the earliest IFU galaxy surveys were unable to explore such correlations in detail as the $R = 850$ spectral resolution of the  Calar Alto Legacy Integral Field Area Survey 
\citep[CALIFA;][]{sanchez12} 
at \Ha\ wavelengths
was too  low to be able to measure velocity dispersions of typical 
current-epoch galactic disks, while the $R \sim 1200$ Atlas3D survey \citep{cappellari11} focused exclusively on early-type galaxies. 
With the current generation of large-scale IFU galaxy surveys such as MaNGA \citep{bundy15} and SAMI \citep{croom12} however
it is now possible
for the first time to study statistically large and representative samples of nearby galaxies with 
spatially resolved spectroscopy.
The MaNGA survey \citep{bundy15}
in particular offers an excellent opportunity to study the dynamical properties of ionized gas across a wide
range of galaxy populations, as it combines an 
order of magnitude increase in sample size compared to previous IFU surveys
with contiguous spectral coverage throughout the $\lambda\lambda 3600 - 10300$ \AA\ wavelength range.

% Outline
In \S \ref{obs.sec} we describe the MaNGA galaxy sample, data reduction process, and the survey data products
used in our analysis.  In \S \ref{defining.sec} we describe the observed relation between the line-of-sight
gas phase velocity dispersion $\sigma_{\Ha}$ and characteristic strong emission line flux ratios, defining a new set
of empirical functions that separate spaxels into dynamically cold and warm populations.
We summarize the physical properties of these populations in \S \ref{properties.sec}, 
noting exceptionally strong correspondence between our new
dynamically-defined selection criteria and a variety of 
other physical observables (e.g., stellar population age,
\Ha\ equivalent width, and radial location) whose relations to strong
line flux ratios have been extensively
discussed in the literature.
We discuss sample-dependent biases in \S \ref{selecteffect.sec}, and
note in particular a significant systematic bias against young, rapidly
star-forming regions in low-mass galaxies in the original SDSS-I sample
compared to the MaNGA IFU sample in \S \ref{sdss1.sec}.
In \S \ref{cloudy.sec} we discuss our results in the context of theoretical photoionization models, and extend our analysis in \S \ref{3d.sec} by considering the folding of the photoionization surface in multiple dimensions. We summarize our conclusions in \S \ref{summary.sec}.
Throughout our analysis we adopt a \citet{chabrier03} stellar initial mass function and a $\Lambda$CDM cosmology in which $H_0 = 70$ km s$^{-1}$ Mpc$^{-1}$, $\Omega_m = 0.27$, and $\Omega_{\Lambda} = 0.73$.

%%%%%%%%%%%%%%%%%%%%%%%%%%%%%%%%%%%%%%%%%%%

\section{Observational Data}
\label{obs.sec}

The SDSS-IV \citep[][]{blanton17} Mapping Nearby Galaxies at APO (MaNGA) galaxy survey has been described thoroughly in the literature via a series of technical papers.
In brief, MaNGA uses custom IFU fiber bundles \citep{drory15} in conjunction with the BOSS spectrographs
\citep{smee13} on the Sloan 2.5m telescope \citep{gunn06} to obtain spatially resolved spectroscopy
at spectral resolution $R \sim 2000$ throughout the wavelength range $\lambda\lambda 3600 -10300$ \AA.
The MaNGA observing strategy and first-year survey data characteristics are described by
\citep{law15} and \citep{yan16b} respectively.
MaNGA data processing occurs in two stages; a Data Reduction Pipeline (DRP) that produces science-grade calibrated spectroscopic data cubes from the raw observational data \citep{law16, law20, yan16a} and a data analysis pipeline (DAP) which uses \textsc{ppxf} \citep{cappellari17} to produce maps of derived astrophysical quantities from the calibrated data cubes \citep{westfall19,belfiore19}.  The MARVIN data interface framework provides a responsive python-based toolkit for interacting with the MaNGA data products both via a web interface and programmatic queries \citep{cherinka19}.

As discussed by \cite{wake17}, the MaNGA galaxy sample is chosen to span both the red sequence and the star-forming main sequence and follow a nearly flat stellar mass distribution in the range $M_{\ast} = 10^9 - 10^{11} M_{\odot}$ .  These galaxies are broken into a primary sample covering galaxies out to 1.5 effective radii
($1.5 R_{\rm e}$), a secondary sample covering out to $2.5 R_{\rm e}$, and a color-enhanced sample designed to more extensively populate
lower-density regions of color magnitude space (e.g., the `green valley').  These samples are selected without regard
to galaxy morphology or the presence of active galactic nuclei, although a variety of smaller ancillary samples within the MaNGA survey ($\sim 15$\% of the total number of data cubes) explicitly target unique galaxy classes such as major mergers, edge-on disks, AGN, dwarf galaxies, etc.

\begin{figure}
\epsscale{1.2}
\plotone{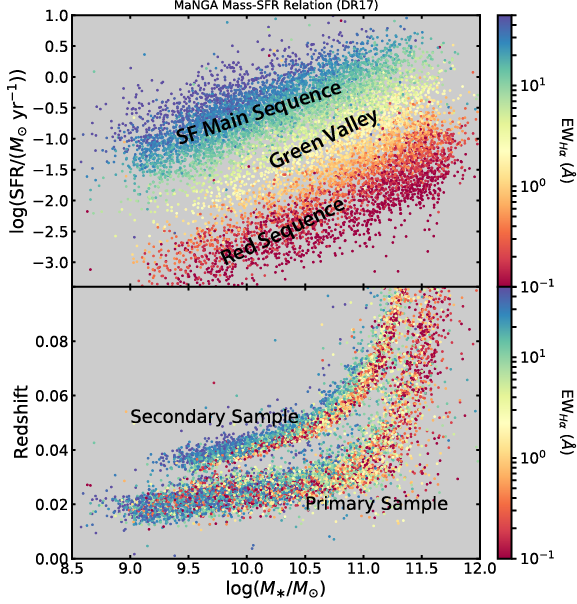}
\caption{Stellar mass vs redshift and total \Ha-derived SFR (uncorrected for dust extinction) 
within the IFU field of view for the MaNGA MPL-11 galaxy
sample.  Each point represents one of the $\sim 10,000$ galaxies in MPL-11, and are color coded by their integrated \Ha\ equivalent width.  The Primary ($1.5 R_{\rm e}$) and
Secondary ($2.5 R_{\rm e}$) samples define a clearly offset trend in redshift at fixed stellar mass.
}
\label{sample.fig}
\end{figure}

For the present analysis, 
we use data products from the final MaNGA internal data release version MPL-11.  
These data products are identical to those that will be released publicly in SDSS Data Release
DR17, and similar to those in MPL-10 discussed by \citet{law20} except for a larger
total sample size
\citep[see summary of all previous versions provided by][their Table 1]{law20}.
Starting with the 11273 galaxy data cubes in MPL-11, we downselect to the 10296 cubes that are not part of ancillary programs targeting small regions within the M31 or IC342 galaxies, or that target intracluster light, ultrafaint dwarfs, or other unusual objects in the Coma cluster \citep[see, e.g.,][]{gu18}.
We additionally discard 158 data cubes flagged by the DRP/DAP as having significant reduction problems, 119 data cubes for which the galaxy was deliberately miscentered in the IFU by more than 1 arcsec (as part of various ancillary programs), and 3 data cubes for galaxies at redshift $z$ below 0.001 to ensure a more consistent range of redshifts for the sample. This leaves a sample of 10016 data cubes, of which 9883 are unique galaxies and a small number have 1-2 repeat observations for cross-calibration purposes.  As illustrated in Figure \ref{sample.fig} these galaxies span three orders of magnitude in stellar mass, and four orders of magnitude in total star formation rate.\footnote{Effective radii and stellar masses are drawn from the parent galaxy catalog
described by \citet{wake17} based on an extension of the NASA-Sloan Atlas
\citep[NSA,][]{blanton11}.  Star formation rates are estimated from the integrated \Ha\ flux in the MaNGA data.}

%The MPL-10 data products used here are broadly similar to previous versions available publicly
%through SDSS Data Releases DR13 \citep{dr13}, DR14 \citep{dr14}, and DR15 \citep{dr15} but differ in both the total number of galaxies and
%the pipeline-provided assessment of the instrumental spectral line 
%spread function.\footnote{Comparable data products for the entire MaNGA galaxy sample will be publicly released in 2021
%as part of SDSS DR17, as will the raw data and all previous versions of the data reduction and analysis pipelines \citep[see summary by][]{law20}.}
At $R \sim 2000$ the MaNGA hardware delivers a $1\sigma$ instrumental velocity resolution of about
70 \kms, posing a substantial challenge for the accurate recovery of astrophysical velocity dispersions
$\sim 15-30$ \kms\ for galactic thin disks, especially since the instrumental resolution is known to vary
with both time and the gravitational flexure of the spectrographs.  As discussed at length by \citet{law20} however,
the improvements in calibration in MPL-10 enable exquisite characterization of the
instrumental response, allowing us to push scientific analyses down to velocity dispersions that are typically
prohibitively difficult to measure for comparable resolution instruments.
Based on comparisons against high-resolution ($R \sim 10,000$) external observations, \citet{law20}
find that the MPL-10 time-dependent LSF provided by the DRP is accurate to 0.6\% (systematic) with 2\% random uncertainty,
and that the resulting
H$\alpha$ velocity dispersions provided by the MaNGA DAP have an uncertainty of about 1 \kms\ (systematic)
and 5 \kms\ (random) at $\sigma_{\Ha} = 20$ \kms\ for lines detected at SNR $> 50$.  Repeating the analyses of \citet{law20}, we find results that are statistically unchanged for MPL-11.

In our present analysis we use the \Ha\ velocity dispersion maps produced by the MaNGA DAP \citep{westfall19, belfiore19} using the `hybrid' binning scheme in which stellar continuum measurements are made on galaxy data cubes that have been voronoi binned \citep{cappellari03} to a minimum
$g$-band continuum SNR of 10 \AA$^{-1}$, while emission line measurements are made for each individual 0.5 arcsec spaxel in
the data cubes.  
The stellar continuum templates used for fitting the emission lines (see further discussion in
\S \ref{contsub.sec}) are based on a series
of hierarchically clustered template spectra observed by the MaNGA MaStar program \citep{yan19}.
We reject individual spaxel measurements with bad data quality flags set for either the \Ha\ flux, velocity, or velocity dispersion, and likewise reject $\sim 100$ spaxels in the MPL-11 sample
with DAP-reported \Ha\ flux in the range
$10^{-13} - 10^{-5}$ erg s$^{-1}$ cm$^{-2}$ spaxel $^{-1}$
%(corresponding to $\sim 1 M_{\odot}$/yr/kpc2 at the median redshift of the MaNGA sample) 
as these represent unmasked noise artifacts.

Additionally, we correct the velocity dispersion maps provided by the DAP for beam smearing following the method
described by \citet{law20}.  In brief, this method involves estimating the magnitude of beam smearing by
convolving a model velocity field with the known point spread function and subtracting the artificial component
of the velocity dispersion in quadrature from the observed values.  The magnitude of
this correction varies depending on the velocity field of a given galaxy, but for
$\sigma_{\Ha} < 40$ \kms\ the median correction due to beam smearing is 9\%, with
90\% of spaxels having a correction of less than 25\%.

%%%%%%%%%%%%%%%%%%%%%%%%%%%%%%%%%%%%%%%%%%%

\section{Defining Dynamical Samples}
\label{defining.sec}

\begin{figure*}
\epsscale{1.15}
\plotone{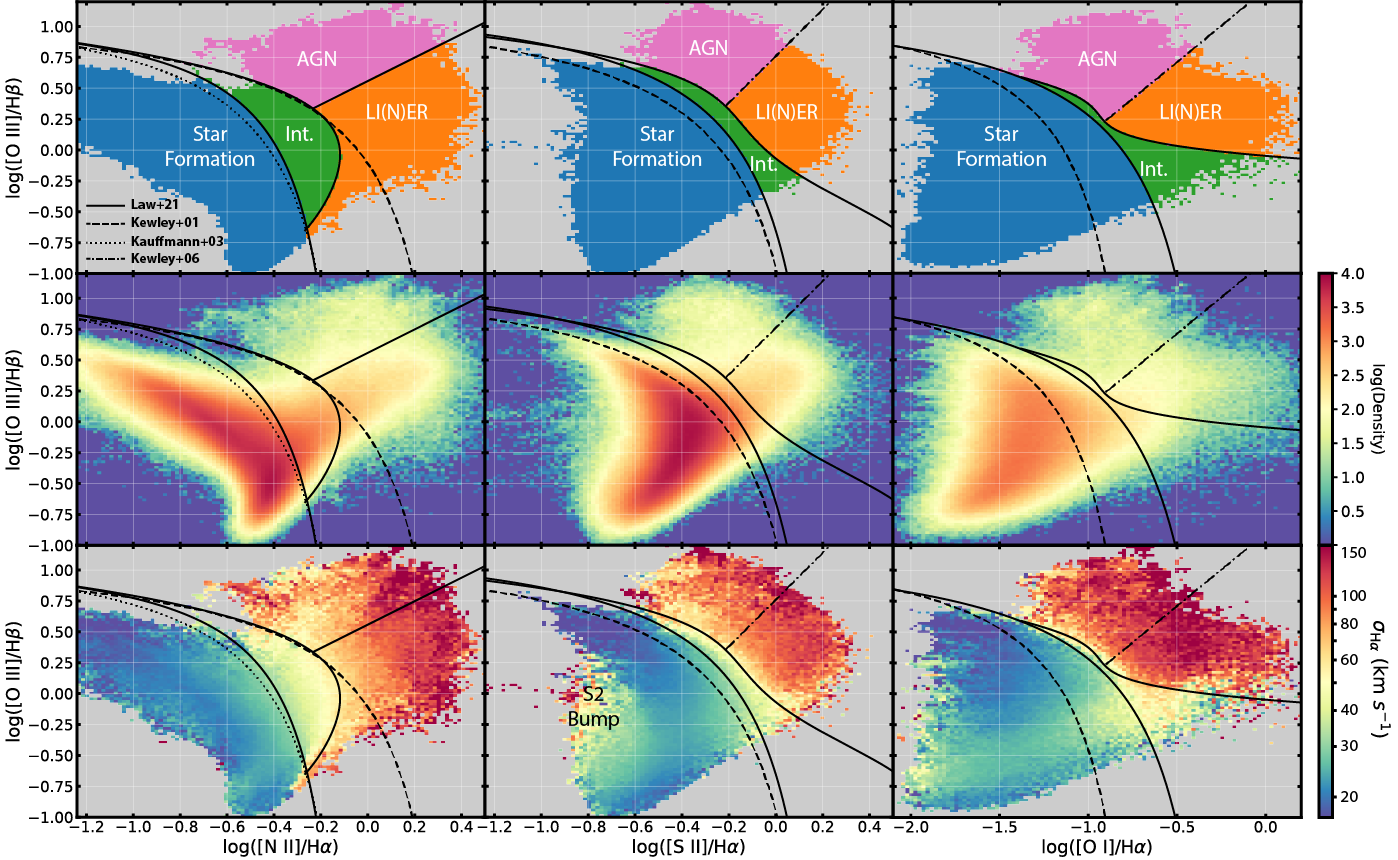}
\caption{
Top panels: Classification of the \ntwo/\Ha, \stwo/\Ha, and \oone/\Ha\ diagnostic diagrams into four regions
(star-forming, intermediate, AGN, and LI(N)ER)
based on their observed dynamical properties.
Middle panels: Number density of spaxels with a given
diagnostic nebular emission line ratio for all 
spaxels in MPL-11 whose
emission lines are detected with SNR $>$ 5 ($\sim 3.6$ million spaxels from $7400$ individual galaxies).  The star-forming sequence is clearly
defined as the red peaks in the density distribution, while lower-density
regions trace emission line ratios typically associated with
active galactic nuclei or LI(N)ER emission.
Bottom panels: As above, but color-coded by the sigma-clipped mean \Ha\ velocity dispersion
of the MaNGA spaxels at a given line ratio.
In all panels the black solid curves represent empirical polynomial fits tracing
the $\sigma_{\Ha} = 35$ \kms\ and 57 \kms\ ionized gas isodispersion contours
(see discussion in \S \ref{defining.sec}), while the black dashed
lines represent the theoretical maximum-starburst relations defined by \citet{kewley01}.
Black dot-dashed lines in the \stwo/\Ha\ and \oone/\Ha\ panels (middle and right columns respectively) show the 
empirical density-based Seyfert/LI(N)ER classifications defined by \citet{kewley06}, while the corresponding black solid
line in the \ntwo/\Ha\ panel (left columns) is defined in Equation \ref{n2agn.eqn}.  The black dotted line in the \ntwo/\Ha\ panels
represents the empirical boundary to the star-forming sequence
from SDSS single-fiber observations defined by \citet{k03}.
%\redtxt{MAB: I'd recommend putting the kinematic classification boundaries in the 3rd or second row with the current, or bottom row at the top -- the distribution of the kinematics is the key result upon which the classification depends.}
}
\label{bpt1.fig}
\end{figure*}

Using the maps of strong emission line fluxes produced by the DAP, we plot in Figure \ref{bpt1.fig} (middle panels)
the density distribution of MaNGA spaxels for a 100 $\times$ 100 pixel grid in the classical line ratio diagnostics
$N2 =$ log(\ntwo\ $\lambda 6583$/\Ha), $S2 =$ log(\stwo\ $\lambda 6716+6731$/\Ha), and $O1 =$ log(\oone\ $\lambda 6300$/\Ha) vs $R3 =$ log(\othree\ $\lambda 5007$/\Hb).\footnote{Throughout this manuscript we follow the use in which \ntwo\ refers specifically to the \ntwo\ $\lambda 6583$ intensity, while \stwo\ refers to the sum of the \stwo\ $\lambda 6716$ and $\lambda 6731$ intensities.  Likewise, $N2, S2, O1$, and $R3$ refer to the logarithmic line intensity ratios, while \ntwo/\Ha, \stwo/\Ha, and \oone/\Ha\ refer to the relevant
diagnostic diagrams.}
In each panel we have selected all spaxels from the galaxy sample for which the relevant emission
lines have all been detected with SNR $> 5$ and whose \Ha\ velocity dispersions are greater than zero after correcting for both beam smearing and the instrumental LSF.\footnote{Rejection of spaxels with zero or imaginary velocity dispersions after quadrature subtraction of
the instrumental LSF and beam smearing rejects just 5\% and 2\% of the total sample respectively.}
For the \ntwo/\Ha\ and \stwo/\Ha\ diagrams this results in a sample size of about 3.6 million spaxels\footnote{Since individual spaxels are 0.5 arcsec in size and the MaNGA PSF has a FWHM of about 2.5 arcsec there is thus significant correlation between adjacent spaxels in a given galaxy.  Conservatively, this corresponds to $\sim$ 300,000 statistically independent spaxels, although in practice the number is higher because not all spaxels in a given resolution element meet our SNR threshold.} distributed across 7400 galaxies, while the \oone/\Ha\ diagram contains about a factor of 3 fewer total spaxels (1.2 million spaxels across 6300 galaxies) because \oone\ is generally fainter
than either \ntwo\ or \stwo.
These samples are subject to significant selection effects, the biases of which we address
in detail in \S \ref{selecteffect.sec}.

As discussed at length by \citet{belfiore16}, 
the spatially resolved SDSS/MaNGA data 
in Figure \ref{bpt1.fig} show similar 
trends as those observed in earlier-generation SDSS single-fiber spectroscopy 
\citep[e.g.,][]{k03, bman04, tremonti04}.
The well-defined sequence at the lower left
corresponds to ionizing photons emitted by young massive stars in galactic HII regions, while
the extended regions toward the top right of the diagram correspond to gas illuminated by photons
with a harder spectrum emitted by a combination of AGN, dynamical shocks, and LI(N)ER-like sources.
The traditional star-forming sequence in the \ntwo/\Ha\ diagram is bounded by the empirical 
number density relation
given by \citet[][black dotted line in Figure \ref{bpt1.fig}]{k03}.
Another commonly-used set of relations are those given by \citet[][black dashed lines in Figure \ref{bpt1.fig}]{kewley01},
who used photoionization models to compute a series of theoretical `maximum starburst' relations in each of the
\ntwo/\Ha, \stwo/\Ha, and \oone/\Ha\ projections.

For each bin in the \ntwo/\Ha, \stwo/\Ha, and \oone/\Ha\ diagrams containing more than 5 spaxels, we compute the sigma-clipped mean
\Ha\ velocity dispersion $\sigma_{\Ha}$, and plot the resulting trends in Figure \ref{bpt1.fig} (lower panels).\footnote{FITS representations of the density and mean \Ha\
velocity dispersion plots can be found in the online supplementary data.}
We immediately note that the traditional star-forming sequence 
(as defined by either the \ntwo/\Ha\ relation of \citet{k03} or the \stwo/\Ha\ relation of \citet{kewley01})
is extremely well defined,
with median gas-phase velocity dispersion $\sigma_{\Ha} < 40$ \kms\ throughout a wide range of line ratios.
 At higher N2, S2, or O1 line ratios
%indicative of a harder ionizing spectrum 
the velocity dispersion rises rapidly, reaching values of 100-200 \kms\
in regions typically associated with AGN or LI(N)ER emission.
Given this strong correlation between the
{\it sources} of the ionizing photons and the line of sight velocity distribution (LOSVD) of the 
gas {\it illuminated} by these photons, we therefore
explore whether it is possible to use our kinematic data (which is broadly representative of the
low redshift galaxy population) to define physically distinct subsamples that can help constrain
theoretical starburst models.

\begin{figure}
\epsscale{1.1}
\plotone{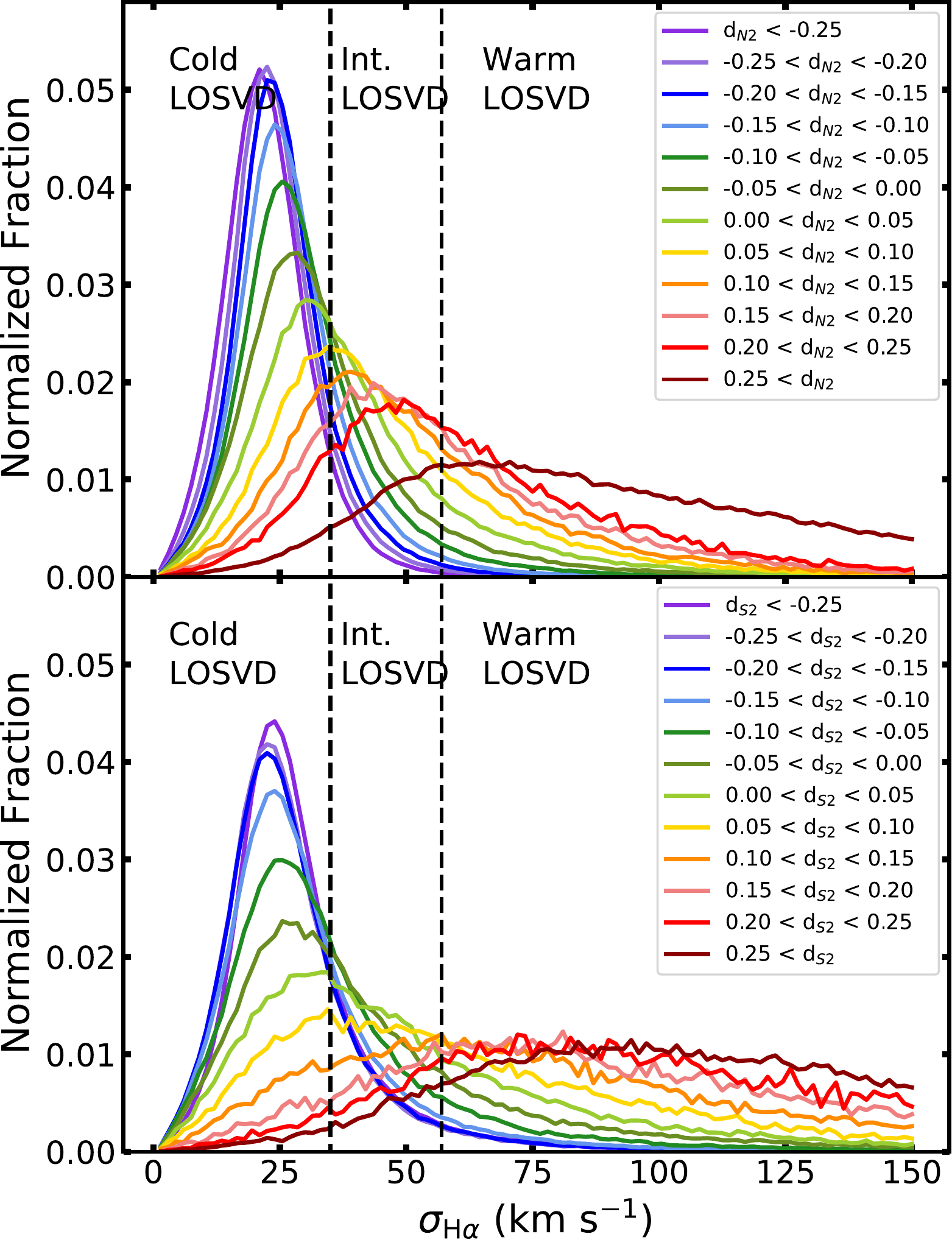}
\caption{Top panel: Histogram of individual spaxel velocity dispersions binned according to their effective distance $d_{N2}$ (measured in dex)
from the \citet{k03} classification line in the
space of emission line intensity ratios \ntwo/\Ha\ vs \othree/\Hb.  
Bottom panel: As above, but for spaxels binned according to their distance $d_{S2}$ from our kinematically-defined
classification line in \stwo/\Ha\ vs \othree/\Hb\ space.
In both panels
the vertical dashed lines denotes the $1\sigma$ and $3\sigma$ wings of the cold gas LOSVD; we 
use these divisions at $\sigma_{\Ha} = 35$ \kms\ and 57 \kms\ to split spaxels into dynamically cold,
intermediate, and dynamically warm populations.
}
\label{bpt_divhist.fig}
\end{figure}

Starting from the \ntwo/\Ha\ relation shown in Figure \ref{bpt1.fig} (left-hand panels), we note
that the \citet{k03} relation does a good job of visually 
tracing the isodispersion contours separating low-$\sigma_{\Ha}$ regions (blue/green points) from higher $\sigma_{\Ha}$ regions (yellow/red points).
The average quantities shown here though belie the even more significant differences
between the underlying {\it distributions} of gas dispersions.
We therefore compute the effective distance $d_{N2}$ of a given spaxel
in $N2$ vs $R3$ space from the \citet{k03} relation, with negative
values defined to represent line ratios below the relation and positive values above the relation.\footnote{Using the shortest distance from the curve rather than the distance along the
vertical axis; see, e.g., discussion in \S 4 of \citet[][]{kewley06}.}
As we illustrate in Figure \ref{bpt_divhist.fig} (top panel), 
the overall trend in the median $\sigma_{\Ha}$
from the star-forming to non star-forming regions is
driven not by a gradual shift in the underlying gas kinematics, but rather by an abrupt
change in the relative weights of two kinematically distinct LOSVDs.  
At $d_{N2} \leq -0.1$ dex spaxels exhibit
a remarkably constant range of velocity dispersions that change only minimally with increasing
distance from the \citet{k03} relation.  This dynamically cold LOSVD is 
strongly peaked around $\sigma_{\Ha} = 24$ \kms\ and has a narrow distribution with 
half-width about 11 \kms.
In contrast, at values of $d_{N2} > 0$ dex, 
$\sigma_{\Ha}$ is increasingly dominated by an entirely different and
much broader LOSVD ranging from  $\sigma_{\Ha} > 30 - 150$ \kms.
These results are reminiscent of the nearly-bimodal distributions observed by \citet{rich11}, and similarly
motivate us to separate the sample into two distinct 
dynamical populations.

As discussed further in \S \ref{selecteffect.sec}, any effort to define a single
`transition' threshold between dynamically cold and warm LOSVDs is necessarily subjective
and dependent on the details of both the sample selection and numerical analysis.
We have chosen to use a $2.5\sigma$ clipping algorithm for instance to identify the
peak of the cold LOSVD, as this method appeared to best reproduce the visual
peak shown in Figure \ref{bpt_divhist.fig} in the presence of extended asymmetric tails
to the distribution.  If we had instead chosen to use
a simple median, a gaussian-fitting technique, or a different
kind of sigma clipping our derived centroid would vary only very slightly in the range
$\sigma_{\Ha} = 23-26$ \kms.  
We opt not to use a flux-weighted mean as this would effectively bias our sample
towards small radii instead of allowing us to compare properties throughout the IFU
field of view (see discussion in \S \ref{sdss1.sec}).
In contrast, we used a $5\sigma$ clipping algorithm
to measure the cold LOSVD profile width in order to better capture the extended tail
of the distribution.  Using a more or less aggressive approach can change the measured
half-width in the range 9-16 \kms.  This width will be driven in part by 
the significant
uncertainty in MaNGA velocity dispersions far below the instrumental limit.  Monte Carlo
tests suggest that these uncertainties may contribute about 6 \kms\ in quadrature (consistent with
Figure 15 of \citet{law20} for our range of SNR), leaving the
astrophysical width in the range 7-15 \kms.

Given the limitations of the MaNGA data this far below the instrumental resolution,
for our present purposes we simply define the $1\sigma$ and $3\sigma$ extremes of the well-defined
cold LOSVD at $\sigma_{\Ha} = 35$ and 57 \kms\ respectively
as the approximate transition points to an `intermediate' and a dynamically `warm' gas distribution.
Using these thresholds, we define an entirely dynamically-based
series of classifications for the  \ntwo/\Ha, \stwo/\Ha, and \oone/\Ha\ diagnostic diagrams.
For the $1\sigma$ lines defining our nominal boundary of the dynamically cold
sequence in Figure \ref{bpt1.fig} we fit the isodispersion contours using
rectangular hyperbolae of the form defined by
\citet{kewley01} and find:

\begin{equation}
\label{1sig_n2.eqn}
    %\textrm{log} \left ( \frac{\othree}{\Hb} \right ) = \frac{0.359}{\textrm{log} (\ntwo/\Ha) + 0.032} + 1.083
    R3 = \frac{0.438}{N2 + 0.023} + 1.222
\end{equation}

\begin{equation}
\label{1sig_s2.eqn}
%    \textrm{log} \left ( \frac{\othree}{\Hb} \right ) = \frac{0.410}{\textrm{log} (\stwo/\Ha) - 0.198} + 1.164
    R3 = \frac{0.648}{S2 - 0.324} + 1.349
\end{equation}

\begin{equation}
\label{1sig_o1.eqn}
%    \textrm{log} \left ( \frac{\othree}{\Hb} \right ) = \frac{0.612}{\textrm{log} (\oone/\Ha) + 0.360} + 1.179
    R3 = \frac{0.884}{O1 + 0.124} + 1.291
\end{equation}

while for the $3\sigma$ lines defining the boundary of the warm sequence we use
a fourth-order polynomial as a function of $R3$ in order to satisfactorily
fit the wide range in observed behaviors and find:

\begin{equation}
\label{3sig_n2.eqn}
N2 = -0.390 \, R3^4 -0.582 \, R3^3 -0.637 \, R3^2 -0.048 \, R3 -0.119
\end{equation}
for $-0.65 < R3 < 1.00$,

\begin{equation}
\label{3sig_s2.eqn}
S2 = -1.107 \, R3^4 -0.489 \, R3^3 +0.580 \, R3^2 -0.579 \, R3 -0.043
\end{equation}
for $-1.10 < R3 < 1.10$, and

\begin{equation}
\label{3sig_o1.eqn}
O1 = 19.021 \, R3^4 -36.452 \, R3^3 + 21.741 \, R3^2 -5.821 \, R3 -0.328
\end{equation}
for $-0.25 < R3 < 0.60$.

The resulting fits are shown in Figure \ref{bpt1.fig} as black solid lines.
Repeating our earlier analysis of the N2 LOSVD histograms as a function of distance from the \citet{k03}
relation using instead the offset distance from our newly-derived 35 \kms\ isodispersion \stwo/\Ha\ relation
(Figure \ref{bpt_divhist.fig}, lower panel) we find an identical $1\sigma$ threshold of
$\sigma_{\Ha} = 35$ \kms\ for the cold LOSVD, confirming that our analysis has converged.

For the \ntwo/\Ha\ diagram (Figure \ref{bpt1.fig}, left-hand column) we note that our $1\sigma$ dynamical relation (Eqn. \ref{1sig_n2.eqn}) 
is extremely similar to the density-based
\citet{k03} relation, albeit protruding to values of N2 that are larger by about
0.05 dex.
Our $3\sigma$ relation (Eqn. \ref{3sig_n2.eqn} 
extends to larger values of N2 and initially follows the \citet{kewley01}
maximum starburst relation at large R3 before looping back
to lower values of N2 at low R3
to accommodate a tail of high $\sigma_{\Ha}$ spaxels.
In contrast, the maximal-starburst line overshoots the
dynamically-cold population, and extends well into the 
range of $\sigma_{\Ha} = 50-100$ \kms\ that are more consistent with large-scale gas flows or other components that are much thicker than traditional
star-forming disks.  This finding is in agreement with recent photoionization models such as those by \cite{byler17} and Belfiore et al. in prep, which more closely match both
our results and the \citet{k03} line, as shown in Sec. \ref{cloudy.sec}.

For the \stwo/\Ha\ diagram (Figures \ref{bpt1.fig} and \ref{bpt2.fig}, middle column)
our $1\sigma$ relation (Eqn. \ref{1sig_s2.eqn}) is similar to the  \citet{kewley01} maximal-starburst line, 
albeit shifted to larger S2 by about 0.15 dex.
Our $3\sigma$ relation (Eqn. \ref{3sig_s2.eqn}) in contrast extends nearly linearly towards the bottom right of the diagram, encompassing
a narrower range of `intermediate' population spaxels than in the \ntwo/\Ha\ diagram.

For the \oone/\Ha\ diagram (Figures \ref{bpt1.fig} and \ref{bpt2.fig}, right-hand column) our $1\sigma$
dynamical relation (Eqn. \ref{1sig_o1.eqn}) 
extends much higher (nearly 0.3 dex) in O1 than the \citet{kewley01}
relation, although the exact track is somewhat uncertain due to the cluster
of higher $\sigma_{\Ha}$ values towards low values of R3.
Our $3\sigma$ relation (Eqn. \ref{3sig_o1.eqn}) 
differs dramatically from the shape of the \ntwo/\Ha\ or \stwo/\Ha\ cases,
extending nearly horizontally along the line R3 $= 0.0$ to encompass a wide `intermediate' region at low values of R3.

All of these relations are subject to our somewhat arbitrary choice of how
to define the boundaries between the dynamically cold, intermediate, and warm kinematic populations.  However, our overall conclusions are largely insensitive to such details.
If we instead repeat our analysis using a slightly different peak velocity dispersion
for the cold LOSVD profile, or using a different profile width, the $1\sigma$ sequences
defined by Eqns. \ref{1sig_n2.eqn} - \ref{1sig_o1.eqn} change by 0.05 dex or less.
Our $3\sigma$ sequences defined by Eqns. \ref{3sig_n2.eqn} - \ref{3sig_o1.eqn}
change by as much as 0.08 dex in the N2-R3 diagram, and less in the S2-R3 diagram.

From a dynamical point of view, the cleanest sample of spaxels with gas-phase velocity dispersions akin
to those of HII regions in a cold disk can thus be obtained using our empirical $1\sigma$ relations in either
\ntwo/\Ha, \stwo/\Ha, or \oone/\Ha\ line ratio space.  In a practical sense, our \ntwo/\Ha\ selection method 
is nearly equivalent to the classical \citet{k03} method, our \stwo/\Ha\ selection method
is broadly similar to the \citet{kewley01} method but includes a slightly larger
range in \stwo/\Ha, and our \oone/\Ha\ selection method differs dramatically from any
previous relations.

\section{Physical Properties of the Dynamical Samples}
\label{properties.sec}

Having defined three dynamically distinct samples (i.e., cold, intermediate, and warm)
in \S \ref{defining.sec}, we next consider how these populations compare in terms of other physical properties.
In doing so, we additionally include one more physical division in the warm population between
the AGN-like and the LI(N)ER-like spaxels.  For the \stwo/\Ha\ and \oone/\Ha\ diagrams we simply adopt the
empirical density-based relations given by \citet{kewley06} in which
\begin{equation}
\label{s2agn.eqn}
    R3 = 1.89 \times S2 + 0.76
\end{equation}
for values of $-0.22 < S2 < 0.3$ (i.e., meeting our $3\sigma$ boundary of the `intermediate' region), and
\begin{equation}
\label{o1agn.eqn}
    R3 = 1.18 \times O1 + 1.30
\end{equation}
for values of $-0.9 < O1 < 0.3$.
Additionally, we define a new and similar relation for the \ntwo/\Ha\ diagram:
\begin{equation}
\label{n2agn.eqn}
    R3 = 0.95 \times N2 + 0.56
\end{equation}
for $-0.24 < N2 < 0.5$ (see discussion in \S \ref{liner.sec}).

\begin{figure*}
\epsscale{1.2}
\plotone{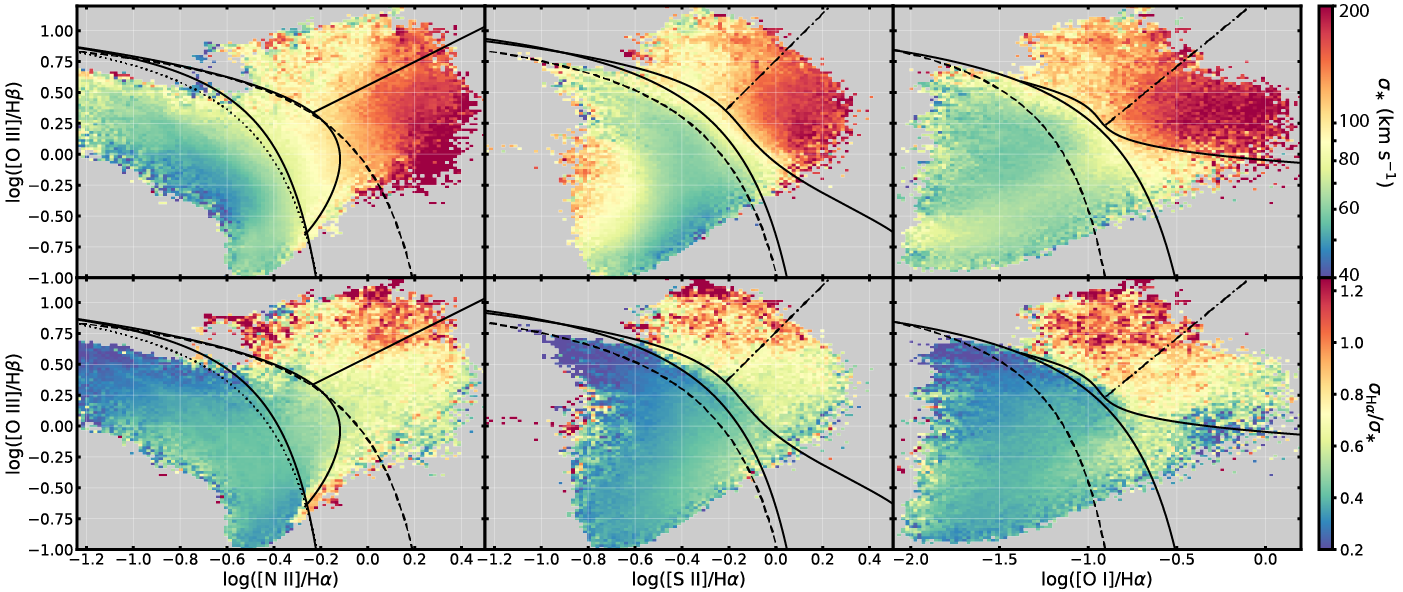}
\caption{As Figure \ref{bpt1.fig}, but showing pixels color-coded by the median
stellar velocity dispersion, and the median ratio between gas-phase and stellar
velocity dispersion. Note the strong evolution of the relative velocity dispersions, in which the ionized gas dispersions are significantly smaller
than the stellar dispersion in the star-forming region, only slightly smaller than the stellar dispersions in the LI(N)ER region, and comparable to or in
excess of the stellar dispersions in the AGN region.
}
\label{bpt2.fig}
\end{figure*}

\begin{figure*}
\epsscale{1.2}
\plotone{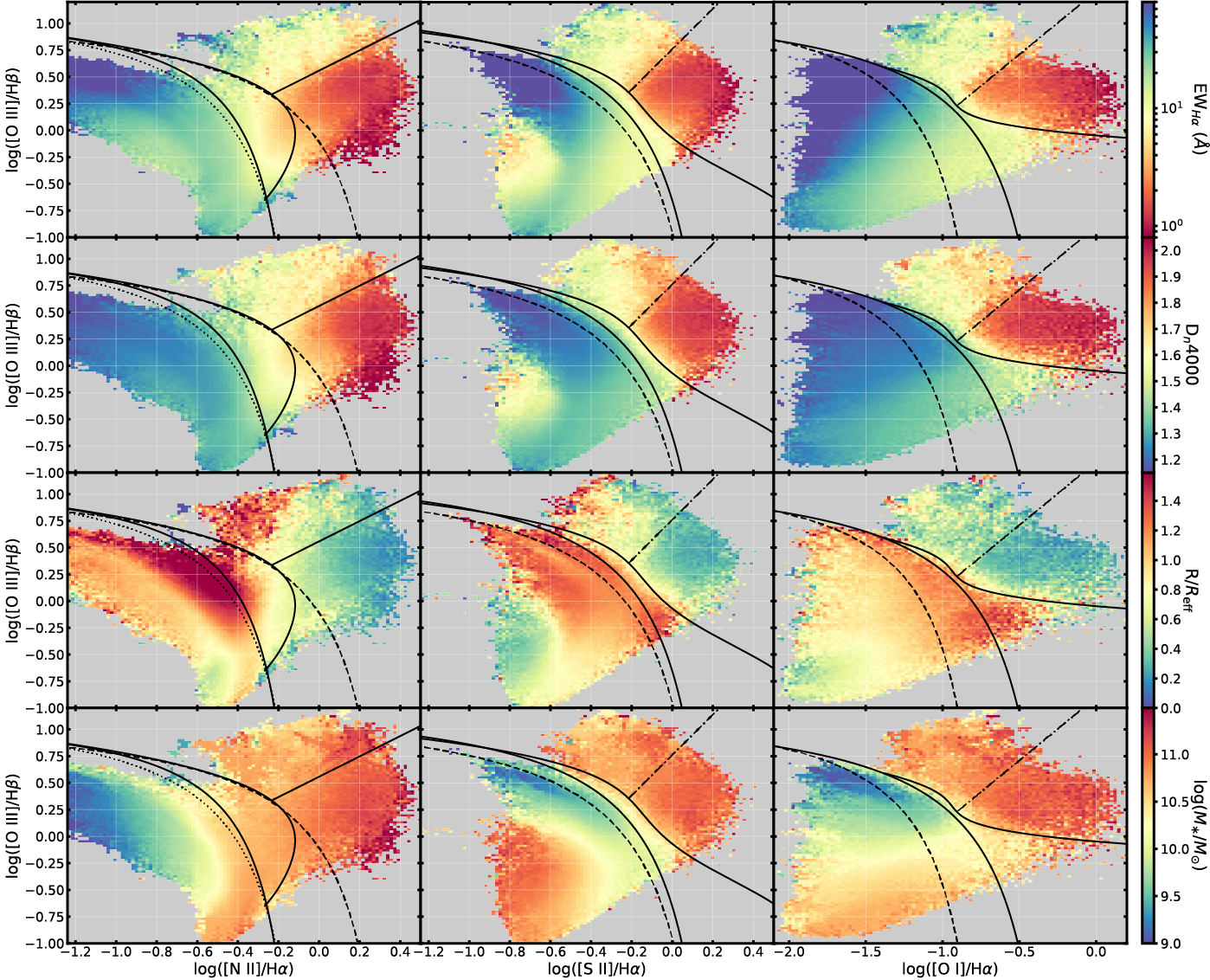}
\caption{As Figure \ref{bpt1.fig}, but showing pixels color-coded by the median
\Ha\ equivalent width,
D$_n$4000 spectral break strength, radius (given as a fraction
of the host galaxy effective radius $R_{\rm e}$), and host galaxy
stellar mass for all MaNGA MPL-10 galaxy spaxels meeting our selection criteria.
Although our classification lines (solid black lines) were defined using
solely dynamical criteria they nonetheless trace clear boundaries in a
variety of other physical observables as well.
}
\label{bpt3.fig}
\end{figure*}

\begin{figure*}
\epsscale{1.1}
\plotone{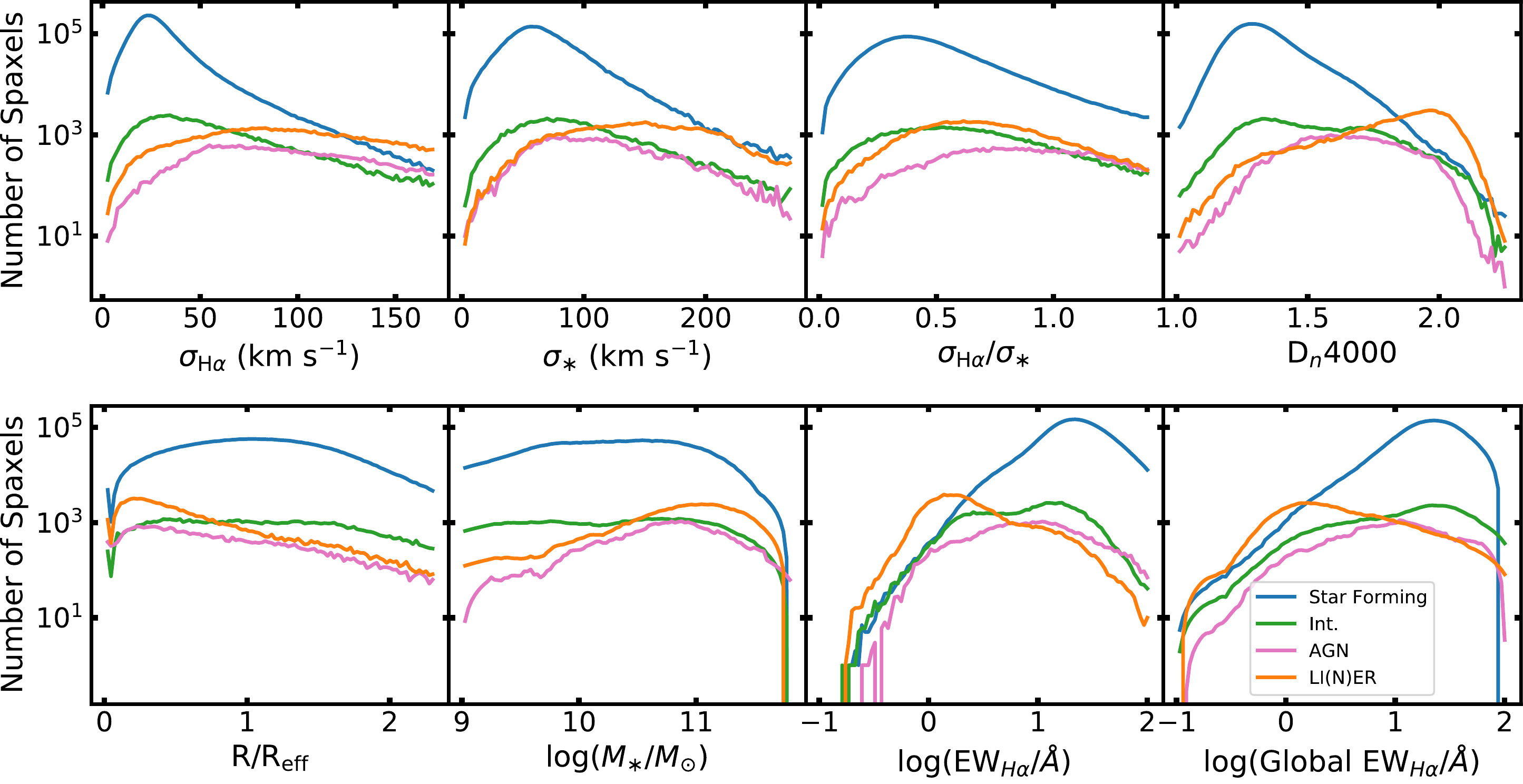}
\caption{Histograms of galaxy properties in each of the four classes
(star-forming, intermediate, AGN, and LI(N)ER) in the \stwo/\Ha\ diagnostic diagram defined by
Figure \ref{bpt1.fig}.  Individual panels show
distributions both in the properties of individual spaxels and in the overall galaxy-integrated
stellar mass and global equivalent width of the host galaxies
within which individual spaxels reside.
Histograms for total galaxy mass and equivalent width have been lightly smoothed by
a Savitzky-Golay filter to minimize high-frequency artifacts from the discrete integrated
values across multiple spaxels.}
\label{histprop.fig}
\end{figure*}

We explore the physical interpretation of the trends observed in Figure \ref{bpt1.fig}
by using the DAP maps in a similar manner to calculate the mean stellar velocity dispersion, the ratio of the stellar and gas-phase velocity dispersion, the \Ha\ equivalent width,
the median strength of the 4000\AA\ break
\citep[D$_{n}$4000, i.e., a proxy for stellar population age,][]{balogh99},
the spaxel radius as a fraction
of the galaxy effective radius ($R/R_{\rm eff}$), and the host galaxy stellar mass.
We illustrate the two-dimensional
distribution of each of these characteristics in the \ntwo/\Ha, \stwo/\Ha, and \oone/\Ha\
diagrams in Figures \ref{bpt2.fig} and \ref{bpt3.fig}, and plot the overall distributions
within each of our four regions in Figure \ref{histprop.fig}.
We note that many of the trends illustrated by Figure \ref{bpt3.fig} in particular are well-established in the literature
\citep[see, e.g.,][and references therein]{sanchez20a}
and provide important context for our new dynamical results.

\subsection{Properties of the Dynamically Cold Sample}

We consider first those galaxy spaxels within the dynamically cold 
population below our 35 \kms\ isodispersion line; these correspond to emission line
ratios within the classical star-formation region.
As discussed in the previous section, not only is the mean gas-phase velocity dispersion
in this region less than in other regions (see, e.g., Figure \ref{bpt1.fig}), but
the overall distribution of values is substantially narrower as well (Figure \ref{histprop.fig}).  The majority of all star-forming spaxels have line of sight
velocity dispersions
$\sigma_{\Ha}$ in the range  $15-35$ \kms\
consistent with a population of HII regions embedded within a dynamically cold gas disk.
These gas-phase dispersions are typically about 40\% of the corresponding stellar
velocity dispersion, which peaks at values of 
$\sigma_{\ast} \sim 50-60$ \kms\ corresponding to a composite thin+thick 
stellar disk. 

Based on Figures \ref{bpt3.fig} and \ref{histprop.fig},
these spaxels unsurprisingly 
\citep[see, e.g.,][their Figures 3 and 1 respectively]{belfiore16,sanchez20a}
have large \Ha\ equivalent
widths EW$_{\Ha} > 10$\AA, and lie almost exclusively within galaxies
whose globally-integrated equivalent widths are typical of
the late-type star-forming galaxy population in Figure \ref{sample.fig}.
Likewise, these spaxels
tend to have a weak 4000 \AA\ break consistent with a stellar population age
of less than $\sim 1$ Gyr for reasonable stellar population models
\citep[see, e.g.,][]{noll09}, and to lie at predominantly large radii $\sim 1 R_{\rm e}$ consistent with inside-out models of galaxy growth.
Such star-forming spaxels are hosted by galaxies across a wide range of stellar
mass $M_{\ast} = 10^9 $ --- $10^{11} M_{\odot}$, for which the nearly-flat
mass distribution is a consequence of the MaNGA sample selection
\citep[see][for details]{wake17}.

As illustrated by Figure \ref{bpt3.fig} (bottom left panel)
the host galaxy mass 
clearly correlates well with the observed N2 line
ratio \citep[i.e., since N2 is a proxy for metallicity the MaNGA spaxels
follow a well-defined mass metallicity relation, see e.g.,][]{tremonti04}.
A similar trend is apparent in the \ntwo/\Ha\ projection of $\sigma_{\Ha}$ as well,
implying a positive correlation such that $\sigma_{\Ha}$ increases 
monotonically from 20-30 \kms\ with increasing stellar mass.
Such correlations between $\sigma_{\Ha}$ and stellar mass 
(or alternatively, with star formation rate)
have been noted previously by other groups
\citep[e.g.,][]{green14,krum18,varidel20}; we defer a detailed discussion of such trends {\it within} the star formation
sequence to a forthcoming publication.

Figures \ref{bpt1.fig}, \ref{bpt2.fig}, and \ref{bpt3.fig} also demonstrate that the three \ntwo/\Ha, \stwo/\Ha, and \oone/\Ha\
diagnostic diagrams are not equivalent in terms of their ability to pick
out a pure sample of star-forming spaxels.  
%Largely-horizontal trends between velocity dispersion and N2 for instance are absent along S2 and O1,
%or tilted such that a larger component of the total discriminatory power lies along the R3 axis instead.
For instance, the \stwo/\Ha\ diagram shows clear evidence of a region
that we term the `S2 bump' within the 
nominally star-forming region around S2 $ = -0.8$ and R3 $ = 0$.  Spaxels with emission line ratios placing them in this region
have abnormally weak \Ha\ equivalent widths, strong 4000 \AA\ breaks,
lie at smaller radii ($< 0.5 R_{\rm e}$) in
higher-mass galaxies ($M_{\ast} > 10^{10.5} M_{\odot}$), and have higher stellar
($\sigma_{\ast} = 80-90$ \kms) and gas-phase ($\sigma_{\Ha} = 50$ \kms) velocity 
dispersions than any spaxels in the N2-defined star-forming region.
A similar feature is visible in the \oone/\Ha\ diagram as well, but manifests more as a 
gradient in each of the above quantities in R3 at fixed O1.

As we demonstrate in \S \ref{3d.sec}, such features are a consequence of the complex folding of the 
multidimensional photoionization hyper-surface, which can produce unusual
artifacts when projected into two dimensions \citep[see, e.g., Fig.~11 of][]{kewley19}.  
The ultimate impact of this unusual group of spaxels
is likely minimal for most observational studies (since the relative number density of 
spaxels in the bump is 2-3 orders of magnitude lower than in the peak
of the star-forming sequence), but is nonetheless of  interest in constraining
numerical models.
We return to this discussion in \S \ref{cloudy.sec} and \ref{3d.sec}.

\subsection{Properties of the LI(N)ER Sequence}
\label{liner.sec}

At the largest S2, N2, or O1 line ratios
stellar populations models \citep[e.g.,][]{kewley01} suggest that the ionizing radiation from
young stars is inadequate to produce the observed strength of
the \ntwo, \stwo, and \oone\ low-ionization emission lines.  
The origin of the photons ionizing the gas in such LI(N)ER regions has thus been debated
in the literature as some combination of faint AGN \citep[e.g.,][]{heckman80,kewley06},
reprocessed leakage from 
HII regions \citep[e.g.,][]{mathis86,dm94,flores11}, radiative shocks \citep[][]{dopita95}, 
or a radially extended population of other
ionizing sources \citep[e.g.,][]{binette94, yan12}.
Most recently numerous authors have favored this latter interpretation given the observed radial surface
brightness profiles of the Balmer emission lines \citep{sarzi10,yan12}, with photoionization from hot evolved post-AGB stars emerging as a leading
candidate \citep[see, e.g.,][]{binette94, sarzi10,singh13, belfiore16, zhang17, byler19}.

As illustrated by Figure \ref{bpt1.fig}, our 3$\sigma$ isodispersion criterion
defined in Eqns. \ref{3sig_n2.eqn} - \ref{3sig_o1.eqn} identifies a set of spaxels with large 
gas-phase velocity dispersions and a range of line ratios consistent with the LI(N)ER galaxy population.
Figures \ref{bpt2.fig} and \ref{bpt3.fig} show that this dynamically-defined boundary also
excellently reproduces
visible boundaries between the LI(N)ER and other regions in terms of physical observables 
such as D$_n$4000, effective radius, stellar velocity dispersion,
and host galaxy mass (particularly for the \stwo/\Ha\ and \oone/\Ha\ projections).
This general agreement across a variety of independent quantities reinforces our 
interpretation of this boundary as one of physical significance.
Similarly, these quantities likewise show significant differences across the 
\stwo/\Ha\ and \oone/\Ha\ AGN/LI(N)ER dividing lines (Equations \ref{s2agn.eqn} and \ref{o1agn.eqn})
proposed by \citet{kewley06} as well.
Although the AGN and LI(N)ER regions are not as well separated in the \ntwo/\Ha\ diagram
as in the \stwo/\Ha\ or \oone/\Ha, since they still show 
broadly similar distinctions we therefore
define a comparable \ntwo/\Ha\ relation in Equation \ref{n2agn.eqn} that can be used
if observations of the \stwo\ or \oone\ lines are not available.

Figure  \ref{histprop.fig} demonstrates that those spaxels
in our dynamically-defined LI(N)ER region
have predominantly low \Ha\ equivalent
widths and strong 4000 \AA\ breaks indicative of a
$\sim 10$ Gyr old stellar population.  Similarly, while they occur primarily in high-mass galaxies whose
galaxy-integrated \Ha\ equivalent widths place them on the red sequence (c.f. Figure \ref{histprop.fig} and Figure \ref{sample.fig}),
there is a tail to the distribution down to lower masses and global equivalent widths $\sim 10$ \AA\ that 
overlaps with the ordinary star-forming galaxy population.
Given that galaxies with small total \Ha\ luminosities have low SNR
it is probable that the apparent lower boundary to the LI(N)ER equivalent
widths in Figure \ref{histprop.fig} is simply an artifact of our SNR $> 5$ spaxel selection criterion, and that observations
deeper than the MaNGA survey spectra
would extend the LI(N)ER sample well into the early-type red sequence population (see \S \ref{contsub.sec} and Fig A1 of \citealt{belfiore16}).

These results are generally consistent with previous observations by (e.g.)
\citet{cf10}, \citet{yan12}, \citet{belfiore16}, and \citet{sanchez20b}.
With the aid of the MaNGA kinematics we can determine (e.g., Figure
\ref{histprop.fig}) that both the
stellar and gas-phase velocity dispersions of our LI(N)ER sample span a wide range
$\sigma_{\ast} = 100-200$ \kms\ and $\sigma_{\Ha} = 50-150$ \kms respectively, with the
median peaked around $\sigma_{\Ha}/\sigma_{\ast} = 0.6$.
These velocity dispersions are inconsistent with gas distributed in a thin disk, and more consistent with
bulges or other largely pressure-supported systems, suggesting that 
our dynamical LI(N)ER sample predominantly traces a diffuse
warm ionized medium with a large vertical scaleheight broadly resembling
that of the stars.  This conclusion is further supported by the
clear correlations between $\sigma_{\Ha}$, spaxel radius, and host galaxy mass in the sense that the highest values of $\sigma_{\Ha} > 150$ \kms\
occur almost exclusively at $R < R_{\rm eff}$ and $M > 10^{10.5} M_{\odot}$.
%for which photoionization by low-mass evolved stars
%distributed throughout a similar volume appears to be a plausible mechanism.
%\redtxt{Awk, rephrase.}

%Notably, given the extremely strong concentration in our LI(N)ER sample 
%towards small effective radii
%there is little signature of the spatially extended `eLIERs' described by \citet{belfiore16}.
%As we demonstrate in \S \ref{intermediate.sec}, this is because the large-radius eLIERs fall instead in
%our Intermediate sample.  Our LI(N)ER sample is thus most representative
%of the original and `classical' LINERs as defined by \citet{heckman80}. 
%[Is this consistent with Belfiore+16?
%answer from FB: I appears I did not study this point in the paper. Only vague hints can be gather from Belfiore+16 of any difference in the BPT diagram positions of the spaxels belonnging to cLIER and eLIER galaxies. For example, see Fig 5 galaxies c and d. However no obvious difference can be gather from Fig 10 of that paper. In any case, I find it fascinating that there should be a difference. This ought to be explained... maybe in future work]

\subsection{Properties of the Intermediate Sequence}
\label{intermediate.sec}

The Intermediate region is defined
as lying between the $1\sigma$ and $3\sigma$ wings of the star-forming population, with mean gas-phase velocity dispersion
between $\sigma_{\Ha} = 35$ \kms\ and 57 \kms.  Figure \ref{bpt1.fig} illustrates
that the shape and size of this region varies considerably; in the \ntwo/\Ha\ diagram it measures
roughly 0.2 dex in width, tapering slightly toward the top of the AGN and bottom of the LI(N)ER
distributions.  In the \stwo/\Ha\ diagram the Intermediate region is significantly narrower in general,
but flares to be wider between the star formation and LI(N)ER sequences than between the
star formation and AGN sequences.  In contrast, in the \oone/\Ha\ diagram the Intermediate
region is extremely narrow between the star-forming and AGN sequences, but the dividing line
becomes nearly horizontal around R3 $= 0.0$, encompassing a range of values
roughly 1 dex wide between the star-forming and LI(N)ER sequences.

In terms of the gas-phase velocity dispersions, individual spaxels in the Intermediate region
range from $\sigma_{\Ha} =  20-100$ \kms, with the majority of spaxels below 50 \kms.
In the \ntwo/\Ha\ diagram in particular (Figure \ref{bpt1.fig}, left-hand panels)
the average velocity dispersion increases with
distance from the star-forming sequence.
This result is generally consistent with the velocity dispersion mixing sequence presented 
by \citet{ho14} for a single 
disk galaxy observed by the SAMI survey,
who similarly found that the lowest values of $\sigma_{\Ha}$ fell below the maximum starburst line \citep[see also discussion by][]{kewley01, davies16, da19}.
Indeed, given the kpc-scale resolution of the MaNGA data, we should
anticipate that each spaxel in practice likely contains gas excited by a range of different
ionization mechanisms that would not be perfectly well differentiated.
With the benefit of a 7000 times greater sample size, we can now much more clearly delineate the rapid increase in $\sigma_{\Ha}$
above the maximum starburst line as radiative shocks and older stellar populations
become increasingly contributors to the overall ionization of the gas.

In the \stwo/\Ha\ and \oone/\Ha\ diagrams, we note that use of these line ratios 
(with varying sensitivity to metallicity, 
temperature, ionization parameter, and other
properties) results in little to no Intermediate sequence between the star-forming
and AGN regions however.  Instead, the Intermediate sample is dominated by spaxels at lower R3 between the star-forming and LI(N)ER sequences, and many of the \ntwo/\Ha-identified Intermediate spaxels fall in the nominal star-forming
region for \stwo/\Ha\ and \oone/\Ha.

As illustrated by Figures \ref{bpt3.fig} and \ref{histprop.fig}, Intermediate region
spaxels (selected via the \stwo/\Ha\ diagram) are extremely similar to those in the star-forming population.
In addition to only marginally larger gas and stellar velocity dispersions, they have
similar radial distribution in galaxies
of comparable stellar mass, a similar peak in D$_n$4000
(albeit with a larger tail to old stellar populations),
and occur in galaxies with similar globally-integrated \Ha\ equivalent widths (again with an enhanced tail to the distribution).
The most significant difference between the two populations is that the Intermediate
population is systematically shifted to lower \Ha\ equivalent widths for the individual
spaxels.
%Overall, this paints the picture that Intermediate spaxels occur in the same late-type star-forming galaxies with young stellar populations as do the star-forming spaxels.

Given the significant overlap
in population distribution, properties, and dynamics with the star-forming sample, the evidence
strongly suggests that these Intermediate spaxels
may simply be those regions
within ordinary star-forming disks that are
either experiencing some degree of contributions from AGN/shocks, or are
dominated by older
stellar populations with higher levels of diffuse ionized gas (DIG, aka the
warm ionized medium) rather than gas in young HII regions.
As discussed by \citet{zhang17}, DIG-dominated regions with low star formation rate surface
density (e.g., between spiral arms) can be shifted into this region of parameter
space if hot evolved stars with hard spectral energy distributions contribute significantly to the ionizing photon budget.

Our dynamically-defined Intermediate and LI(N)ER distributions may therefore simply 
represent DIG in actively star-forming and quiescent galaxies respectively, with a continuous gradient
between the two distributions in terms of their stellar population age, mass, and stellar/gaseous velocity dispersion.
In predominantly star-forming galaxies the contribution of DIG-like line emission shifts a given spaxel to larger S2 while having only a marginal effect on the line width since the disk stellar velocity dispersion
is relatively small.  
Indeed, we note that both \citet{bizyaev17} and \citet{levy19} found a $\sim 1-2$ kpc scaleheight for DIG in edge-on galaxies
from the MaNGA and CALIFA surveys respectively, consistent with the observed asymmetric drift relative to the galactic HII regions.
In predominantly quiescent galaxies the DIG-like line emission ratios are more extreme due to less residual contribution from HII
regions, and the line widths are broader reflecting the overall
increase in the importance of pressure support to these evolved stellar systems.
Such a picture is consistent with Figure 4 of \citet{belfiore16}, which similarly found a significant increase in $D_n 4000$ as a function
of distance to higher S2 ratios from the \citet{kewley01} line.

%%%%

\subsection{Properties of the AGN Sequence}

At larger values of R3, the AGN sequence is clearly
distinguishable from both the star-forming and LI(N)ER sequences.  Spaxels in this
region are generally thought to have large forbidden-line fluxes resulting from some
combination of extended narrow-line region (ENLR) gas directly ionized by hard radiation
from the central accretion disk \citep[e.g.,][]{groves04,kewley13} and radiative
shocks resulting from the turbulent collision of feedback-driven gas flows
with the surrounding ISM \citep[e.g.,][]{allen08,rich11}.

As illustrated in Figures \ref{bpt3.fig} and \ref{histprop.fig},
the AGN sequence occupies an intermediate
range of \Ha\ equivalent widths (both in terms of individual spaxels and the galaxies
in which those spaxels reside) and D$_n$4000 spectral break strengths
characteristic of 
intermediate-age galaxies in the green valley and
lower-SFR regions of the blue cloud.
AGN-dominated
spaxels are strongly biased toward smaller radii in higher host galaxy masses though, with typical
radii $\sim 0.25 R_{\rm e}$ (albeit with a tail of the distribution
extending above 1$R_{\rm e}$) and
stellar mass $M_{\odot} \sim 10^{11} M_{\odot}$.  The AGN/shock sequence 
thus has a similar
mass distribution to the LI(N)ER sequence, but slightly younger stellar population ages
and galaxy-integrated line emission more consistent with quenching late-type galaxies
than with early-type galaxies.

In term of their kinematics, AGN-dominated spaxels have
a wide range of gas-phase velocity dispersions ranging from $\sigma_{\Ha} = 50 - 200$ \kms\ or
more.\footnote{The MaNGA DAP is not designed to properly fit the extremely broad
components characteristic of Type I AGN \citep[see Section 11 of][]{westfall19},
and these values are therefore incomplete at the highest values of $\sigma_{\Ha}$.}  
Unlike LI(N)ERs the ratio between the gas-phase and stellar
velocity dispersions in the AGN sample is commonly near unity \citep[see also][]{ilha19},
with a significant number of spaxels exhibiting
gas dispersions up to a factor of two in excess of the stellar velocity dispersion.
At the lower end of this distribution, these velocities are consistent with
ENLR gas directly illuminated by the accretion disk.  At the upper end of the
distribution however the gas is perhaps most likely to be shocked and not
in dynamical equilibrium, as models suggest that the observed velocity
dispersion of such gas should be consistent with the $\sim 100-200$ \kms\ velocities
of the shocks themselves \citep[e.g.,][]{rich11}.
Intriguingly, Figure \ref{bpt3.fig} suggests that AGN/shock-like spaxels at predominantly large vs small effective radii may be separable in terms of
their typical N2 line ratio; however, this may simply reflect 
metallicity variations in the gas \citep[e.g.,][]{groves04,allen08} rather than a distinction in the ionization
source itself.

%%%%%%%%%%%%%%%%%%%%%%%%%%%%%%%%

\section{Sample-Dependent Selection Effects}
\label{selecteffect.sec}

In \S \ref{defining.sec} we defined our four broad populations based on a 
division of the \ntwo/\Ha, \stwo/\Ha, and \oone/\Ha\ diagrams
using the $1\sigma$ and $3\sigma$
limits of the clearly defined dynamically cold star-forming sequence.  This division was
based on the distribution of gas-phase velocity dispersions within the entire MaNGA galaxy sample, which is designed to be
broadly representative of the $z=0$ galaxy population \citep[see][]{wake17}.  However, as we explore here there are nonetheless a variety
of both astrophysical and observational selection
effects that are important to consider both within the MaNGA sample itself
and when comparing to other samples.

\begin{figure}
\epsscale{1.1}
\plotone{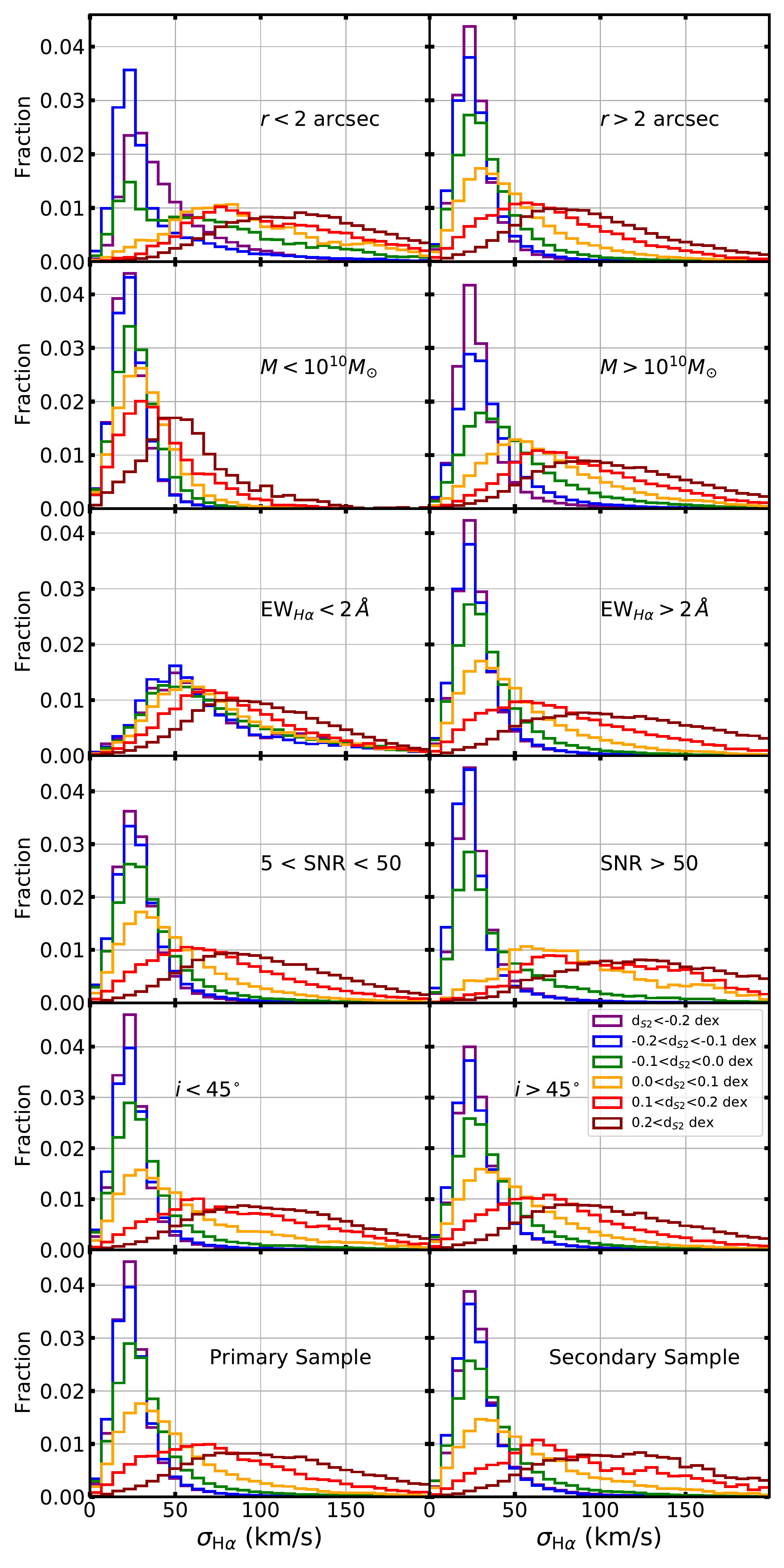}
\caption{As Figure \ref{bpt_divhist.fig}, but binned according
to distance from our best-fit classification line in the \stwo/\Ha\
diagram, and split according to a variety of physical
and observational parameters.
}
\label{bpt_hist.fig}
\end{figure}

\subsection{Observational Selection Effects}

In Figure \ref{bpt_hist.fig} we repeat our earlier exercise from Figure \ref{bpt_divhist.fig},
but focusing on the \stwo/\Ha\ diagram and dividing the MaNGA
spaxels into subsamples according to a variety of physical properties.

\begin{figure}
\epsscale{1.2}
\plotone{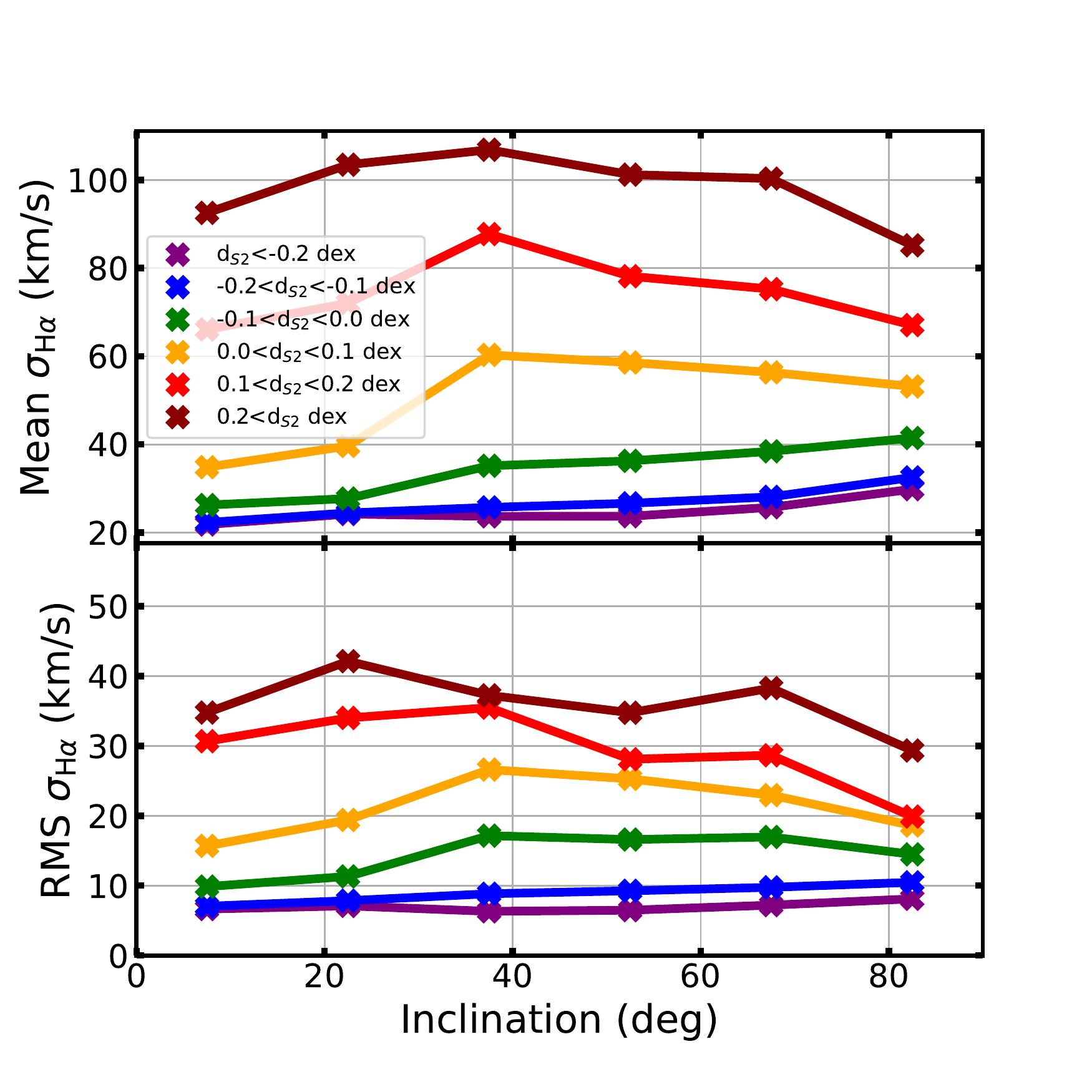}
\caption{$2\sigma$-clipped mean and RMS velocity dispersion as a function of inclination
for various bins of distance $d_{S2}$ from
our kinematically-defined classification line in the \stwo/\Ha\ vs \othree/\Hb\ space.
Edge-on galaxies (90$^{\circ}$ inclination) show marginally larger velocity dispersions than
face-on galaxies (0$^{\circ}$ inclination).
}
\label{inclination.fig}
\end{figure}

To start, we note a few properties that {\it do not} affect our conclusions.
First, galaxy inclination (as derived from the NSA catalog) has only a marginal 
impact on the observed velocity dispersion trends,
as subsamples with $i < 45^{\circ}$ and $i > 45^{\circ}$ both produce statistically
identical distributions.  If we break the galaxy sample into a greater number of
bins (Figure \ref{inclination.fig}) we note a marginal increase in velocity dispersion
for star-forming spaxels in the most edge-on systems, consistent with the superposition
of distinct velocity components along the line of sight.  While this increase will be important
for the study of individual galaxies, it has only a minor impact on the MaNGA ensemble result
as 80\% of our selected spaxels with $d_{S2} < 0$ lie in galaxies with $i < 60^{\circ}$.

Second, the physical resolution of the MaNGA data does
not appear to significantly affect our conclusions either.    At the typical 2.5 arcsec
angular resolution of MaNGA, the effective spatial resolution of the data is about 2
kpc at the median redshift of the sample and individual resolution elements will
therefore be composed of gas ionized by a mixture of physical processes (e.g., AGN, HII regions, 
interarm diffuse ionized gas, etc.).  We therefore compare results between the
MaNGA Primary and Secondary samples\footnote{A simple redshift cut would impose
a significant differential in stellar mass as well given the increase in stellar
mass with redshift in the MaNGA sample (see Figure \ref{sample.fig}).}, 
which at median redshifts of 0.025 and 0.046
correspond to physical resolutions 1.3 and 2.3 kpc respectively.
Other than a small increase in the peak of the cold LOSVD from 24 to 27 \kms\
(possibly due to an increase in the intra-spaxel velocity gradient; see further
discussion in Law et al., in prep) there is no significant difference between the two, suggesting
that our overall results are not a strongly varying function of spatial resolution.

In contrast, spaxels with higher SNR show a much sharper transition between entirely-cold and entirely-warm LOSVDs (i.e., a minimal Intermediate sequence)
compared to spaxels
in which \Ha\ emission was detected at lower SNR.
For the SNR $< 50$ subsample
the transition between populations occurs gradually over 
an Intermediate population of width 0.2-0.3 dex, while the 
SNR $> 50$ subsample jumps between an entirely-cold LOSVD and an 
entirely-warm LOSVD  between 
adjacent bins with $-0.1 < d_{S2} < 0.0$ and $0.0 < d_{S2} < 0.1$ dex.

One possible explanation for this difference may be the larger uncertainties and
known systematic bias of the low-SNR tail of the MaNGA sample to higher $\sigma_{\Ha}$ due to the 
exclusion of spaxels whose observational uncertainties scatter them below the instrumental
resolution \citep[see discussion by][their section 4.3]{law20}.  However, we do not believe
this to be the primary explanation since even low-$\sigma_{\Ha}$ spaxels in our low-SNR subsample in Figure \ref{bpt_hist.fig}
have a median \Ha\ SNR $= 36$, for which uncertainties 
and biases are not substantially different than for the high-SNR
subsample \citep[][their Figure 15]{law20}.
Likewise, in the regions of parameter space where the median SNR is low (i.e., median SNR $\sim$ 20 in the traditional LI(N)ER region) the intrinsic velocity dispersion is high and thus systematic biases in measured $\sigma_{\Ha}$ are negligible.
Unsurprisingly, the primary effect of reconstructing Figure \ref{bpt1.fig} using a more stringent threshold of SNR $> 50$ is thus to severely restrict
the number of measurements in regions with weak overall \Ha\ emission.

Rather, we believe the difference in the high and low SNR subsamples in Figure \ref{bpt_hist.fig}
to be due to the correlation between SNR and spaxel radius, in the sense that applying
a more-aggressive SNR $> 50$ cut systematically biases the observational sample to smaller radii,
effectively downweighting the Intermediate sample compared to the AGN/LI(N)ER samples.
Indeed, the sigma-clipped mean spaxel radius in the Intermediate region is 0.5 $R_{\rm e}$ when
using an SNR $> 50$ cut, compared to 1.1 $R_{\rm e}$ when using an SNR $> 5$ cut
(along with shifting to substantially older ages and higher masses).
Surveys that are shallower than MaNGA (or that sample a different range of
radii; see \S \ref{sdss1.sec}) would thus derive a different width for the Intermediate
region, as may surveys that are substantially deeper.

\subsection{Astrophysical Selection Effects}

In terms of total galaxy mass, we note an appreciable difference
between the low-mass and high-mass galaxy subsamples.
As illustrated by Figure \ref{bpt_hist.fig} (middle panels), galaxies of all masses
show a similar dynamically cold LOSVD component corresponding to HII regions
in the galactic thin disk.
At low masses $M_{\ast} < 10^{10} M_{\odot}$, even spaxels
in the AGN and LI(N)ER regions exhibit $\sigma_{\Ha} < 100$ \kms.  Likewise,
spaxels in such low-mass galaxies are correspondingly skewed to younger ages
and larger \Ha\ equivalent widths, with
stellar velocity dispersion profiles peaked around $\sigma_{\ast} = 60$ \kms.
At larger masses $M_{\ast} > 10^{10} M_{\odot}$ however, the upper bound of the warm 
LOSVD gas increases to $\sim 200$ \kms\ for the LI(N)ER and AGN samples, along with a corresponding
increase in stellar population age and stellar velocity dispersion.
This result is consistent with the deeper gravitational potential wells
of larger mass galaxies, but more likely corresponds simply to the increasing prevalence of
green-valley and red-sequence/early-type galaxies in the upper mass end of the galaxy distribution.
While the relatively flat mass distribution of the MaNGA sample in the range
$M_{\ast} = 10^9 - 10^{11} M_{\odot}$ (see, e.g., Figure \ref{histprop.fig})
has more high-mass galaxies than a purely volume-limited sample \citep[see, e.g.,][]{weigel16}, the addition of more low-mass galaxies
would simply accentuate the dominance of the star-forming sample in an analysis
of spaxels with detectable nebular line emission.

By far the most pronounced difference in gas LOSVDs is observed however if we
compare galaxies whose \Ha\ equivalent width is greater or less than 2\AA\
(which are found within the blue cloud and red sequence respectively,
see Figure \ref{sample.fig}).  
As illustrated by Figure \ref{bpt_hist.fig}, the gaseous LOSVD of 
spaxels with EW$_{\Ha} > 2$ \AA\ broadly resembles the master spaxel sample
with both cold and warm components.  In contrast, there is no dynamically cold
component to the LOSVD in spaxels selected with EW$_{\Ha} < 2$ \AA; the 25 \kms\
galactic HII region signature is completely absent, even in cases where the emission
line flux ratios are nominally consistent with photoionization by young stars.
Similarly, these spaxels live almost entirely in galaxies with stellar mass
$\sim 10^{11} M_{\odot}$ and old stellar populations with strong 4000 \AA\ breaks.
Indeed, in such spaxels the LI(N)ER and Intermediate regions dominate the distribution, 
containing nearly six times more spaxels than the nominally star-forming region.
The effectiveness of the $EW_{\Ha}$ criterion in thus 
efficiently selecting LI(N)ER type spaxels is in keeping with the prior findings
of \citet{yan06}, \citet{cf10}, \cite{sanchez12a,sanchez20b},
\citet{belfiore16}, and \citet{lacerda18}, who proposed using it as an additional classification axis
beyond the usual \ntwo/\Ha, \stwo/\Ha, and \oone/\Ha\ diagrams.

%%%%%%%%%%%%%%%%%%%%%%%%%%%%%%%%

\subsection{Analytical Selection Effects}
\label{contsub.sec}

As outlined in \S \ref{obs.sec}, we have used the versions of the MaNGA DAP data products that
model the stellar continuum underneath the emission lines using hierarchically clustered sets of
stellar spectra from the MaStar program \citep{yan19}, which used the MaNGA IFUs during bright time
to observe a roughly 30,000 star stellar library using the same hardware and software as used
to observe the MaNGA galaxies.
As discussed in detail by \citet[][see particularly their Figs. 10 and 11]{belfiore19},
the choice of stellar continuum model used can have an impact on the recovered emission line properties
by altering the effective profiles of the Balmer absorption lines atop which emission is superposed.

In order to assess the impact of this template choice on our results, 
we repeated our analysis using a second set of
DAP data products that instead used theoretical SSP models based upon the MaStar library
\citet{maraston20} that have systematically deeper Balmer line profiles.  In the dynamically cold star-forming sequence we find that the difference between
the hierarchically clustered stellar spectra and the SSP models is inconsequential, with the SSP-derived
velocities systematically larger by just 0.5 \kms\ (i.e., well within observational uncertainty).
Unsurprisingly, the largest difference occurs in the LI(N)ER region of the emission line diagnostic diagram
which is dominated by extremely low \Ha\ equivalent widths.  At fixed SNR selection criteria,
the LI(N)ER region is significantly more well populated using the SSP models compared to the hierarchically clustered templates, adding about 150,000 new spaxels to those shown in Figure \ref{bpt1.fig}.  These spaxels are located predominantly in high-mass ($M_{\ast} > 10^{11} M_{\odot}$) red sequence galaxies
and add about 700 such galaxies to the overall sample with emission line regions meeting our SNR criteria.
The addition of these extra spaxels does not change our overall conclusions, but increases the median \Ha\ velocity dispersion in the LI(N)ER region from 93 to 100 \kms.

%%%%%%%%%%%%%%%%%%%%%%%%%%%%%%%%

\subsection{Key Differences from the SDSS-I Sample}
\label{sdss1.sec}

The radius-dependent selection effects described in the previous sections have
significant implications for the comparison of our results to those
based on the earlier-generation SDSS single fiber spectroscopy
\citep[e.g.,][]{k03, tremonti04, kewley06, stasin06}.  By virtue of the 3 arcsec diameter of the original SDSS-I spectroscopic fibers, 
that sample was subject to a severe radial selection bias compared
to the integral-field SDSS-IV/MaNGA observations.

The \citet{k03}, \citet{tremonti04}, and \citet{stasin06} samples for instance consisted of 
about  123000, 53000, and 20000 galaxy spectra respectively with a median
redshift $z \approx 0.1$ at which the SDSS fiber subtended a radius of about 3 kpc.
Similarly, the \citet{kewley06} sample contained 85000 galaxies
in the redshift range $z = 0.04 - 0.1$, for which the SDSS fiber subtended a radius
of 1-3 kpc.  In contrast, the SDSS-IV MaNGA sample is both significantly larger
($\sim $ 300000 statistically independent spectral samples with $> 5\sigma$ emission line
detections; see \S \ref{defining.sec}) and covers a much larger radial extent
of 9 kpc on average at the median redshift $z = 0.04$.

\begin{figure*}
\epsscale{1.1}
\plotone{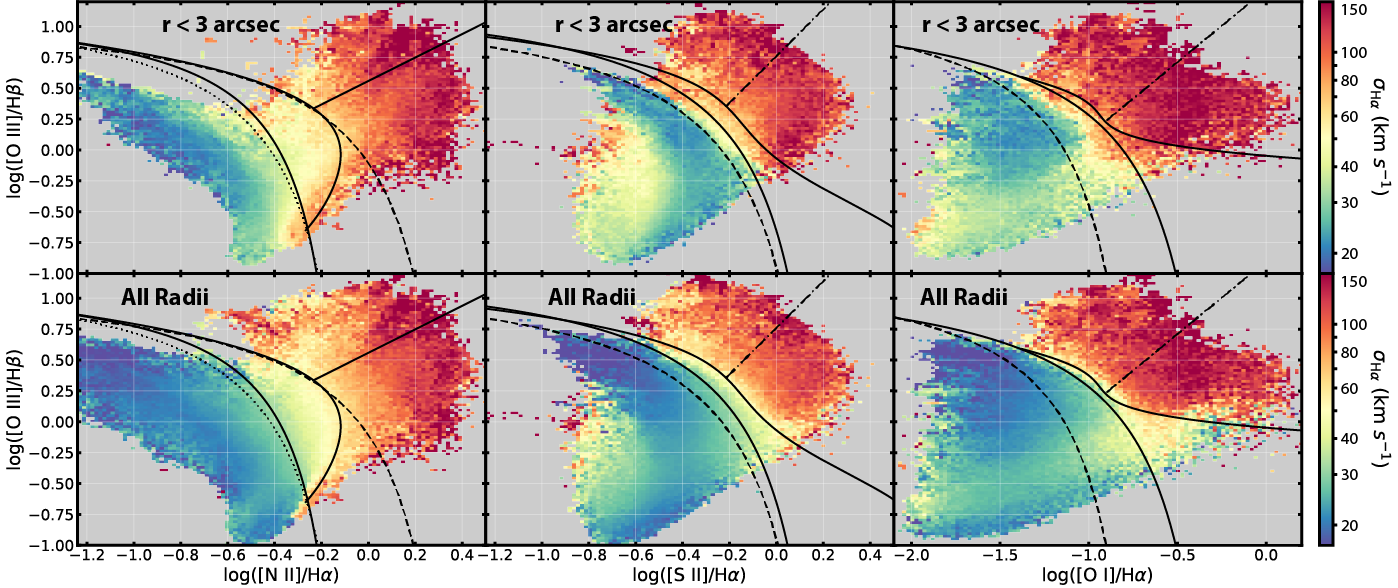}
\caption{As Figure \ref{bpt1.fig} (lower panels), but comparing the results from
the entire MaNGA spaxel sample (lower panels) against a subsample of spaxels
from the central regions of galaxies chosen to closely mimic the selection effects
of the original SDSS-1 spectroscopic survey (upper panels).  Note the shrinking
of the Intermediate region in the low-$r$ subsample, along with a general shifting
to lower line ratios of the warm LOSVD component.
}
\label{bpt_rcompare.fig}
\end{figure*}

We assess the impact of this bias on the SDSS-I sample by
applying an $r < 3$ arcsec cut to the MaNGA data.\footnote{$r < 3$ arcsec at the median
redshift of the MaNGA sample corresponds nearly to $r < 1.5$ arcsec at $z=0.1$; the
\citet{kewley06} sample radial cut is thus even more restrictive than we model here.}
As illustrated in Figure \ref{bpt_rcompare.fig}
the median gas velocity dispersion as a function of strong line intensity ratios
changes enormously with such a radial cut.  Compared to the full MaNGA sample,
the restricted central radial range loses a large fraction of the cold disk sequence,
shrinks the Intermediate population to produce a much sharper
division between cold and warm LOSVDs, 
and shifts the warm LOSVD population towards the lower left in the \ntwo/\Ha, \stwo/\Ha, and \oone/\Ha\
diagrams.  For instance, we note that while the \stwo/\Ha\ and \oone/\Ha\ relations of
\citet[][dashed line in Figure \ref{bpt_rcompare.fig}, middle and right panels]{kewley01}
underpredict the range of the cold LOSVD population when considering all
MaNGA spaxels, they do a much better job of matching the dynamical dividing
line when considering only spaxels at $r < 3$ arcsec.

\begin{figure*}
\epsscale{1.2}
\plotone{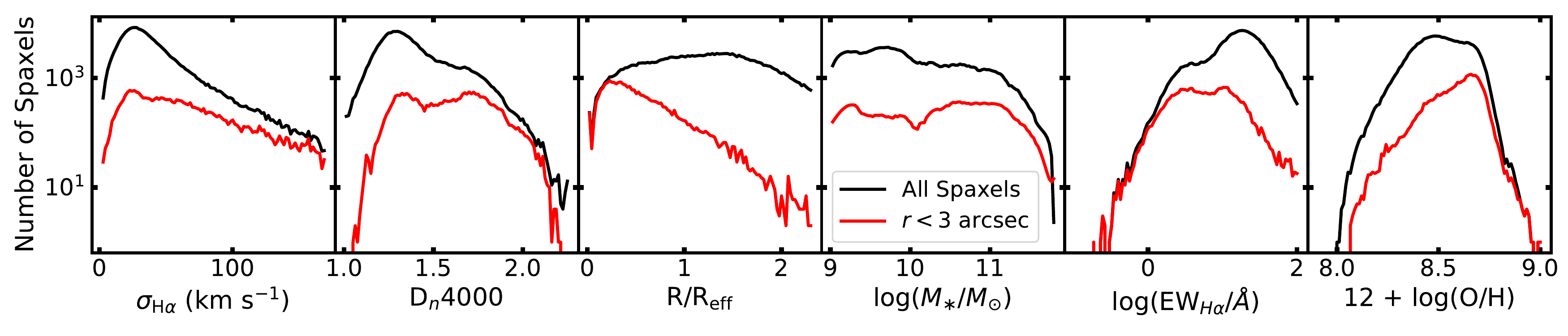}
\caption{Histograms of the physical properties of spaxels with S2 and
R3 line ratios placing them above the maximal starburst relation of
\citet{kewley01} but below the dynamical $1\sigma$ relation defined in Eqn. \ref{1sig_s2.eqn}.
Black and red histograms respectively illustrate the distribution for spaxels
at all radii vs spaxels with $r < 3$ arcsec comparable to the regions that would
have been contained within the original SDSS-I spectroscopic fiber apertures.
The $r < 3$ arcsec sample misses 93\% of the stars in the parent sample, preferentially
those with young ages and high \Ha\ equivalent widths in low mass galaxies.}
\label{smallr_hist.fig}
\end{figure*}

As we show in Figure \ref{smallr_hist.fig} this is because
the SDSS-I radial selection function
misses a statistically important sample of high \Ha\ equivalent width spaxels
in regions with young stellar populations.
In the full MaNGA sample 
these spaxels dominate the region in the \stwo/\Ha\ and \oone/\Ha\ diagrams 
above the \citet{kewley01} relation but below
our dynamical 1$\sigma$ relation given in Eqn. \ref{1sig_s2.eqn}.  Since they are
primarily located at larger radii $\sim 0.5 - 2 R_{\rm e}$ these spaxels are
almost entirely absent from the SDSS-I sample.

Given the strong radial concentration of the AGN and LI(N)ER samples
(Figure \ref{bpt_hist.fig})
a radial selection bias at the same time artificially increases the numerical significance
of these two populations relative to the star-forming and intermediate samples.
As illustrated by Figure \ref{smallr_hist.fig} an SDSS-I-like radial cut includes
nearly all of the high-D$_n$4000, low-EW$_{\Ha}$, high-$\sigma_{\Ha}$ spaxels in the full MaNGA sample, while
undercounting the low-D$_n$4000, high-EW$_{\Ha}$, low-$\sigma_{\Ha}$ spaxels by nearly two orders of magnitude.\footnote{\citet{belfiore16} similarly remarked upon this effect,
noting both the extension above the \citet{kewley01} demarcation line in the \stwo/\Ha\ BPT diagram of MaNGA star-forming spaxels and the virtual absence of the large population of intermediate-D$_n$4000 LI(N)ER-like spaxels in SDSS.}

These differences are not limited to the narrow region above the \stwo/\Ha\ line
of \citet{kewley01}.  Using the $O3N2$ metallicity relation defined by \citet{pettini04}:
\begin{equation}
12 + {\rm log}(O/H) = 8.73 - 0.32 \times (R3-N2)
\end{equation}
we find that the full MaNGA spaxel sample in the dynamically cold $1\sigma$ region
is more metal-poor than the $r < 3$ arcsec
subsample of the MaNGA data.  The shift in the median value from 
$12 + {\rm log}(O/H) = 8.69$ to 8.63 dex is driven by the increased prevalence of
spaxels with $12 + {\rm log}(O/H) \sim 8.5$ dex at large radii
\citep[see also][]{sanchez14}, which has 
significant implications for the
construction of photoionization models (as we discuss further in \S \ref{cloudy.sec}).

In a spatially resolved sense, these differences are clearly of interest
both for what they tell us about the extended structure of galaxies
and for their implications for stellar photoionization models
(see discussion in \S \ref{cloudy.sec}).
In an integrated sense however their significance is more ambiguous since
the total galaxy light tends to be dominated by the central regions.
Indeed, if we compute the \Ha\ flux-weighted mean velocity dispersion
in each bin in Figure \ref{bpt_rcompare.fig} instead of the sigma-clipped mean
(which is effectively an area weighting) we find very little difference
between the low-radius and all-radius subsamples since the low-radius points
dominate in both cases.

In either sense however, caution is warranted when applying 
line-ratio selection
criteria derived from central spaxels in the low-redshift universe to 
identify star-forming regions in the distant universe.
Unlike the SDSS-I $z=0$ sample, seeing-limited studies at higher redshifts tend to
include a much larger fraction of the galaxy in the optical slit, all of which
may have stellar populations that are substantially younger and lower
metallicity than typically
found in the central regions of nearby galaxies.  Likewise, 
adaptive-optics fed IFU studies at high-redshift
\citep[e.g.,][and references therein]{law09, fs18} typically
resolve the galaxy emission with comparable kpc-scale resolution to MaNGA.
The offsets 
in median metallicity and the upper envelope of the
star-forming sequence between the low and high-radius subsamples of MaNGA
data are thus potentially significant compared to the 
observed offset in the
mass-metallicity relation between the $z=0$ sample
and galaxy populations at higher redshifts \citep[e.g.,][]{erb06, maiolino08, sanders20}.
Although the majority of this offset appears to be due to
intrinsic physical differences between the galaxy populations
\citep[e.g., a harder stellar ionizing spectrum at 
fixed nebular metallicity due to alpha enhancement,][]{steidel16,strom17,topping20},
the systematic underepresentation of low metallicity regions
in the SDSS single-fiber comparison sample
should not be neglected.

%As we discuss in more detail in a forthcoming contribution though,

%potential limitations 
%of our beam-smearing correction in central regions \citep[see discussion by][]{law20}
%as such errors would tend to smear out any trends
%in $\sigma_{\Ha}$ rather than producing a {\it sharper} transition.

%This distinction between central and outer spaxels may, however, explain 
%some systematic differences described previously by \citet{belfiore16} between spatially-resolved views of diagnostic line diagrams and previous view provided by
%SDSS single-fiber spectroscopy that contains only the central regions of galaxies.
%[More.  What is the physical mechanism?]

%Based on discussion in Zhang+17, try dividing by the HA Lumunosity density, 
%in erg/s/kpc2.  We find there isn't a clear shift in the histograms.
%In part this is because (as per Zhang+17 fig 2), a single value of S2
%covers a large range in HA SB.  Also might get confused because while
%DIG regions have low HASB, AGN regions seem to have high HASB.
%Unclear this is telling us much.

%%%%%%%%%%%%%%%%%%%%%%%%%%%%%%%%%%%%%%%%%%%

\section{Comparison to Theoretical Models}
\label{cloudy.sec}

\begin{figure*}
%\epsscale{1.1}
\plotone{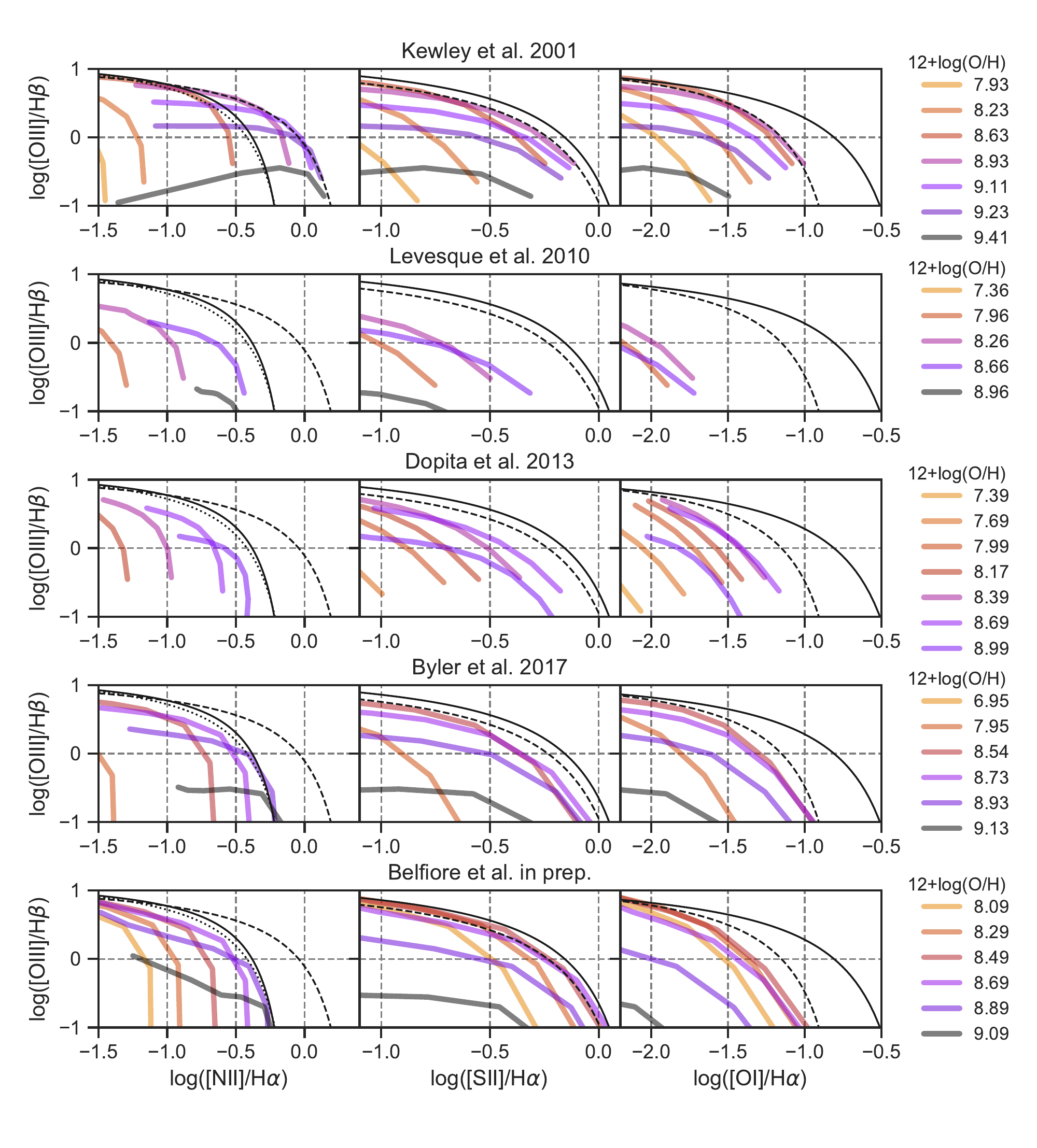}
\caption{Photoionisation models predictions from the literature in the \ntwo/\Ha\, \stwo/\Ha\ and \oone/\Ha\ BPT diagrams. Each row presents sets of models taken from the papers cited in the figure. In each panel, the coloured lines are models computed for a fixed oxygen abundance (given in the legend) and a range of ionisation parameters. LogU generally decreases from the top left to the bottom right in all BPT planes, and models by different authors cover different ranges and value of logU. The black lines represent proposed demarcation lines for star-forming galaxies. The dashed lines are from \cite{kewley01}, and were proposed as a representation of the envelope of their models shown in the topmost row. The dotted line is the empirical demarcation line from \cite{k03} for the \ntwo/\Ha\ diagram. The solid lines are the $1\sigma$ demarcation lines presented in this work (Equations \ref{1sig_n2.eqn} - \ref{1sig_o1.eqn}). The recent models by Belfiore et al., in prep, present an excellent fit to the demarcation lines proposed in this work in the \ntwo/\Ha\ and \stwo/\Ha\ BPT diagrams. No set of models in the literature is capable of reproducing the new proposed demarcation line in the \oone/\Ha\ BPT.
}
\label{bpt_photomodels.fig}
\end{figure*}

Photoionization models have been used by several authors to constrain the position of HII regions (and therefore star-forming galaxies) in the BPT diagram. \cite{kewley01} presented  the first set of easily-applicable demarcation lines in the BPT diagram based on modern ionising continuum models. In detail, \cite{kewley01} present models computed using the Mappings III photoionisation code
\citep[][]{binette85,sd93} and using ionising spectra from two different spectral synthesis codes, Starburst99 
\citep{leitherer99} and Pegase 2 \citep{fr97}. The locations of the HII regions in the BPT diagrams predicted by these two sets of models are remarkably different. \cite{kewley01} argued in favour of the PEGASE 2 models, because of their harder extreme ultra violet (EUV) slope, which provided a better fit to observations of a small sample of starburst galaxies then available. These PEGASE 2 \cite{kewley01} models are shown in the first row of Figure \ref{bpt_photomodels.fig}. Each coloured line represents a set of models with the same metallicty, spanning a range of ionisation parameters, which in the BPT planes decreases going from the top left to the bottom right. The characteristic `folding over' of the models in the BPT planes generates a well-defined envelope for star-forming models, which is fitted by the \cite{kewley01} demarcation lines (dashed black lines in figure). 

The success of the predictions of \cite{kewley01} using the PEGASE 2 models in the \stwo/\Ha\ and \oone/\Ha\ BPT diagrams have been clearly demonstrated by \cite{k03}, who compared them with line ratio measurements from early SDSS fiber spectroscopy. The excellent fit to these observations provided by the \cite{kewley01} demarcation lines in the \stwo/\Ha\ and \oone/\Ha\ BPT diagrams has cemented the use of these boundaries in subsequent works.

The \cite{kewley01} models did not provide an equally good representation of the envelope of star-forming galaxies in the \ntwo/\Ha\ BPT diagram. This prompted \cite{k03} to define an empirical demarcation line in the \ntwo/\Ha\ diagram (black dot-dashed in Fig. \ref{bpt_photomodels.fig}) to replace the \cite{kewley01} predictions. The subsequent denomination of the region between the \cite{kewley01} and \cite{k03} lines in the \ntwo/\Ha\ diagram as `intermediate' has regrettably contributed to the continued use of the incorrect \ntwo/\Ha\ \cite{kewley01} demarcation line, despite the progress made by subsequent authors to build models that more accurately match the empirical \cite{k03} line.

\cite{levesque10} expand on the work of \cite{kewley01} by using STARBURST99 with the newer 
\citet{paul01} and \citet{hm98} atmospheres, which include continuum metal opacities. They compute photoionization models with an updated version of MAPPINGS, which includes a more sophisticated treatment of dust \citep{groves04}. In Fig. \ref{bpt_photomodels.fig} we show the position in the BPT planes of their models calculated using the `high' mass-loss tracks, a constant SFH, a Salpeter IMF with a 100 $M_{\odot}$ upper cutoff, and an age of 6 Myr. Despite the important theoretical improvements, the position of these models is close to that of the STARBURST99 models computed by \cite{kewley01} (not shown), but substantially offset with respect to the PEGASE 2 \cite{kewley01} models. In detail, in the \ntwo/\Ha\ diagram the envelope of the \cite{levesque10} models is shifted to the lower-left, providing a poor fit to the observational data, but in the opposite sense than the  \cite{kewley01} demarcation line. The envelopes in the \stwo/\Ha\ and \oone/\Ha\ diagrams are also offset to the bottom-left of the plane with respect the data, producing an overall much worse fit to both the SDSS single-fiber observations
and the MaNGA results presented in \S \ref{defining.sec}
than the original \cite{kewley01} models.

%\redtxt{Need to rephrase to add more comparison to new results throughout, stress SDSS single fiber and MaNGA differences}

\cite{dopita13} present models with an upgraded version of MAPPINGS (IV), but using the older STARBURST99 atmospheres also employed in \cite{kewley01} (in part to avoid the issues presented in \citealt{levesque10}, who used the newer atmospheres). The \cite{dopita13} models come substantially closer to the envelopes of the observational data in all BPT diagrams. In Fig. \ref{bpt_photomodels.fig} we show the models with a constant SFH, an age of 4 Myr and a Salpeter IMF (and adopting the canonical Maxwell–Boltzmann distribution of electron energies).
While \cite{dopita13} compute models with metallicity 12+log(O/H) $>$ 9.0, such models occupy an unphysical area of the BPT diagram to the far bottom left outside the plot area presented in the figure. 

\cite{byler17} present models based on modern stellar atmospheres for hot stars and simple stellar population models generated by FSPS \citep{conroy09}. In particular, \textsc{fsps} makes use of O and B star spectra generated with WMBasic \citep{paul01}, while Wolf-Rayet stars are taken from the spectral library of \cite{smith02}. Their model simple stellar populations are computed with both Padova \citep{marigo08} and MIST 
\citep{choi16,dotter16} isochrones. All models are computed with the photoionisation code CLOUDY v13.03 \citep{ferland13}. In Fig. \ref{bpt_photomodels.fig} we show the Padova models for a 1 Myr SSP. Considering that this work makes use of modern stellar atmospheres, they have largely improved on \cite{levesque10} by coming substantially closer to all BPT demarcation lines (although code differences, dust depletion factors, and relative abundance prescriptions will also have an effect).
Moreover, unlike the original \cite{kewley01} models they correctly reproduce the \cite{k03} demarcation line in the \ntwo/\Ha\ diagram. The MIST isochrone models (not shown) provide an equally good match, but continue to match the data for older ages, thanks to the inclusion of stellar rotation.

% DRL: I've commented this out for purposes of collaboration review, but kept all of the notes so that we can try to converge
%\redtxt{(Renbin: the discussion above only focuses on the update of the stellar ionizing spectra. Are other factors like abundance pattern all fixed among these papers? These factors are all tied together. For example, using an incorrect abundance prescription could lead to an incorrect choice of stellar spectra. Furthermore, matching the empirical demarcation in 2D does not necessarily mean the models are consistent with the data as they can still be significantly offset in 3D or higher dimensional line ratio space as shown by Ji \& Yan (2020). )}

The most recent update of the CLOUDY code \citep[v17,][]{ferland17}  includes a change in the dielectronic recombination coefficient for $S^{++}$ \citep{badnell15}, which was previously unknown, and estimated by the charge-normalised mean dielectronic recombination rates for the C, N and O \citep{ali91}. Given the importance of the dielectronic recombination channel for $S^{++}$ in nebular conditions, the update caused an increase in the predicted flux of the \stwo\ lines. Belfiore et al. (in prep), recomputed models based on the same formalism as \citep{byler17}, but with this updated CLOUDY version. In the last row of Fig. \ref{bpt_photomodels.fig} we show models computed using the same input physics as \citet{byler17}, MIST isoschrones and age of 2 Myr. The only significant change with respect to the previous \citet{byler17} models
is in the the \stwo/\Ha\ BPT diagram, where the new models lie above the \cite{kewley01} demarcation line.
However, as we demonstrated in Figure \ref{smallr_hist.fig} this is precisely the region in which MaNGA data show evidence
for a substantial population of young, high equivalent-width
and low-metallicity spaxels whose large radii meant that they
were missed by the original SDSS fiber spectroscopy against which the \citet{kewley01} models are traditionally compared.
In contrast, the new models of Belfiore et al. (in prep) almost exactly match our new dynamical boundary to the star-forming
sequence defined in \S \ref{defining.sec} (black solid lines in Figure \ref{bpt_photomodels.fig}).
In summary, modern photoionisation models are capable of reproducing the demarcation lines of \cite{k03} (or the nearly identical kinematically-defined line proposed here) in the \ntwo/\Ha\ diagram,
while the \cite{kewley01} line in this diagram should no longer be used. Recent updates to the atomic data for $S^{++}$ lead to model predictions which follow the \stwo/\Ha\ demarcation line proposed in this work almost exactly.

%Finally, features such as the `S2 bump' may be explained from a theoretical perspective by the folding over of models in the BPT planes. For example, considering the Belfiore at al in prep. models in Fig. \ref{bpt_photomodels.fig} last row, one can see that high-metallicity models will populate the regions of the `S2 bump' at S2 = -0.7 and O3 = -0.5. 

In contrast, no set of models presented in the literature extends to the right of the \cite{kewley01} line in the \oone/\Ha\ diagram,
and the MaNGA spaxels in this range whose gas-phase velocity dispersions are consistent with HII regions in thin disks
are therefore not currently reproducible theoretically.
In part, this may be because 
contamination from diffuse ionised gas can
affect the observed line ratios in low-surface brightness regions probed by MaNGA and contribute (at least partially) to their higher O1. While N2 and S2 are also enhanced in the diffuse ionised gas with respect to HII regions, the effect is of the order of 0.4 dex, while for O1 the DIG show higher line ratios by up to 1.2 dex \citep{zhang17}, therefore constituting a much more severe contaminant.
However, as we demonstrated in \S \ref{sdss1.sec}, the spaxels in this region 
generally have large \Ha\ equivalent widths unlike those observed from DIG.
At the same time, the \oone\ line emission is notoriously difficult to model accurately 
(see discussion by Ji et al., in prep) because it is produced in the narrow partially ionised zone at the boundary between HII regions and neutral clouds, and it is strongly affected by the presence of shocks or diffuse ionised gas. The relative role of young star-forming regions and 
contributions from diffuse ionised gas to line ratios measured on kpc-scales
therefore remains an active area of research and needs to be assessed with higher spatial resolution observations of nearby star-forming galaxies (Belfiore et al., in prep).

Another complicating factor is that the definition of a `best-fit' photoionization model is itself ambiguous. In the discussion above we have assumed that the envelopes of the ideal photoionization models should match exactly the $1\sigma$ dynamical demarcation lines observed in Figure \ref{bpt1.fig} to within some uncertainty associated with variations in secondary model parameters (e.g. stellar SEDs, dust depletion factors, prescriptions for secondary elements, etc.) in the observed H\,{\sc ii} regions. However, as suggested by \cite{ji20}, from a statistical point of view the most representative photoionization model found in a multidimensional line ratio space would not necessarily match the demarcation lines in any given two-dimensional projection of that space. To better evaluate the goodness of fit, it is thus important to check the consistency of model predictions across multiple emission line ratios \citep[see e.g.][]{vogt14, mingozzi20}.

%, which is beyond the scope of this work. As a simple exercise however, we consider the velocity dispersion of ionized gas using three-dimensional diagnostic diagrams in \S\ref{3d.sec}, and demonstrate how the kinematic information in three dimensions is connected to what we see in two dimensions.

%%%%%%%%%%%%%%%%%%%%%%%

\section{Separating different kinematic components in a three-dimensional line-ratio space}
\label{3d.sec}

A consistency check 
reveals that of 3.2 million spaxels that satisfy either our 
\ntwo/\Ha\ or \stwo/\Ha\ star-forming selection criteria\footnote{Applying a
consistent $5\sigma$ SNR requirement on both \ntwo\ and \stwo.},
88\% simultaneously satisfy both selection criteria, 11\% are indicated to be
star-forming using the \stwo\ selection but not the \ntwo\ selection, and 1\%
are indicated to be star-forming using the \ntwo\ selection but not the \stwo\
selection.  The major reason for differences between \ntwo\ and \stwo\ selection
techniques is thus the relatively large fraction of spaxels indicated to be
star-forming by the \stwo\ technique but not the \ntwo, 86\% of which fall
within the Intermediate classification region for the \ntwo\ diagram.
The median velocity dispersion of this latter population is 47 \kms\ (i.e.,
substantially larger than the 24 \kms\ median for spaxels for which the \ntwo\
and \stwo\ techniques agree upon a star-forming classification), suggesting
that the \stwo\ technique incorporates a relatively large fraction of spaxels
that are not dynamically cold.  Features like the `S2 bump' in the S2-R3 diagram are mainly composed of such misclassified spaxels.

If we instead define star-forming spaxels to be those with dynamically cold
\Ha\ velocity dispersions $\sigma_{\Ha} < 35$ \kms, we can calculate both the 
purity (i.e., the fraction of spaxels meeting a given line-ratio selection criterion
that are dynamically cold) and the completeness (i.e., the fraction of dynamically cold
spaxels that meet a given line-ratio selection criterion) of all three \ntwo, \stwo,
and \oone\ selection techniques.\footnote{This approach will be limited by our
appreciable uncertainties on individual spaxel velocity dispersions and by the
questionable validity of our assumption that dynamically cold velocity dispersions
are necessarily and uniquely star-forming.  However, it nonetheless provides
a convenient numerical point of comparison.}
As indicated by Table \ref{purity.table}, the N2-R3 method has better selection
purity than the S2-R3 or O1-R3 methods, but lower completeness.

\begin{deluxetable}{ccc}
\tablecolumns{3}
\tablewidth{0pc}
\tabletypesize{\scriptsize}
\tablecaption{Purity and Completeness of Spaxel Selection Techniques}
\tablehead{
\colhead{Method} &  \colhead{Purity} & \colhead{Completeness}}
\startdata
N2-R3 & 83\% & 95\% \\
S2-R3 & 77\% & 99\% \\
O1-R3 & 75\% & 99\% \\
$P_1$-$P_2$ & 83\% & 96\% 
\enddata
\label{purity.table}
\end{deluxetable}

%As demonstrated in \S\ref{cloudy.sec}, the folding over of the photoionization models in the standard ionization diagnostic diagrams not only indicates that there will be upper envelopes in the data distribution, but also that composite regions (created by the mixing of the spatially close H\,{\sc ii} regions and LINERs or AGN regions) with high enough metallicities could scatter below the demarcation line.
%In principle, it may be possible to resolve these degeneracies in a 
%multidimensional parameter space, thus producing fewer misclassifications.

Recently, \cite{ji20} proposed to use a 3D diagnostic diagram composed of N2, S2, and R3 to obtain a cleaner separation of different ionization mechanisms and found that the photoionization models (which are 2D surfaces with varying metallicity and ionization parameter) that represent H\,{\sc ii} regions and AGN regions are well separated in 
both 3D and in carefully chosen reprojections of this space.
Using velocity dispersion as a proxy for `true' star-forming spaxels, we
explore whether a similar reprojection can better highlight the separation
between star-forming and non star-forming spaxels than the traditional axes.

\begin{figure*}
    %\animategraphics[autoplay,loop,width=0.33\textwidth]{3}{animation/3d_vd_}{0}{35}
    \includegraphics[width=0.33\textwidth]{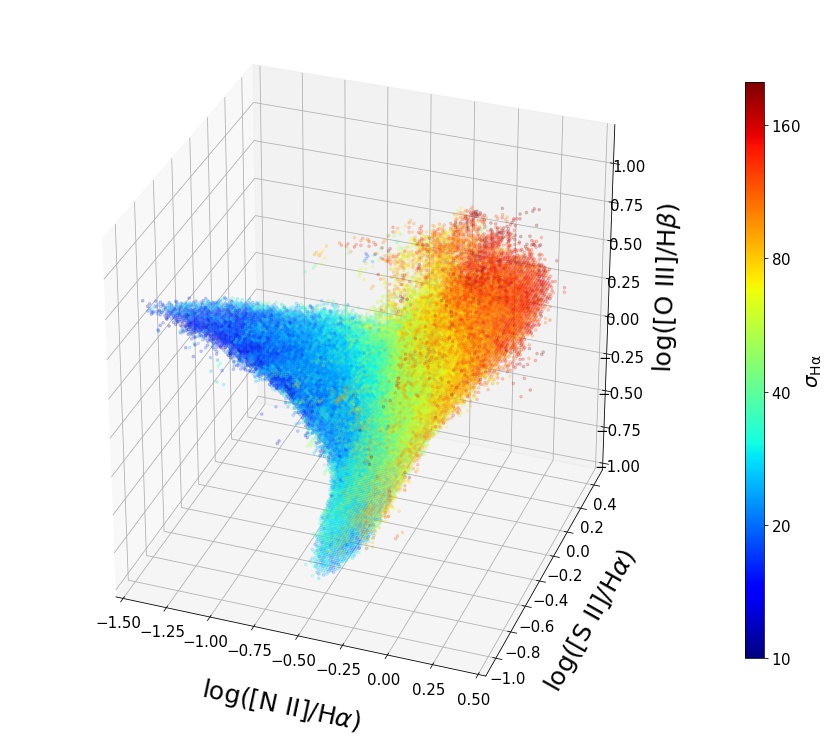}
    \includegraphics[width=0.33\textwidth]{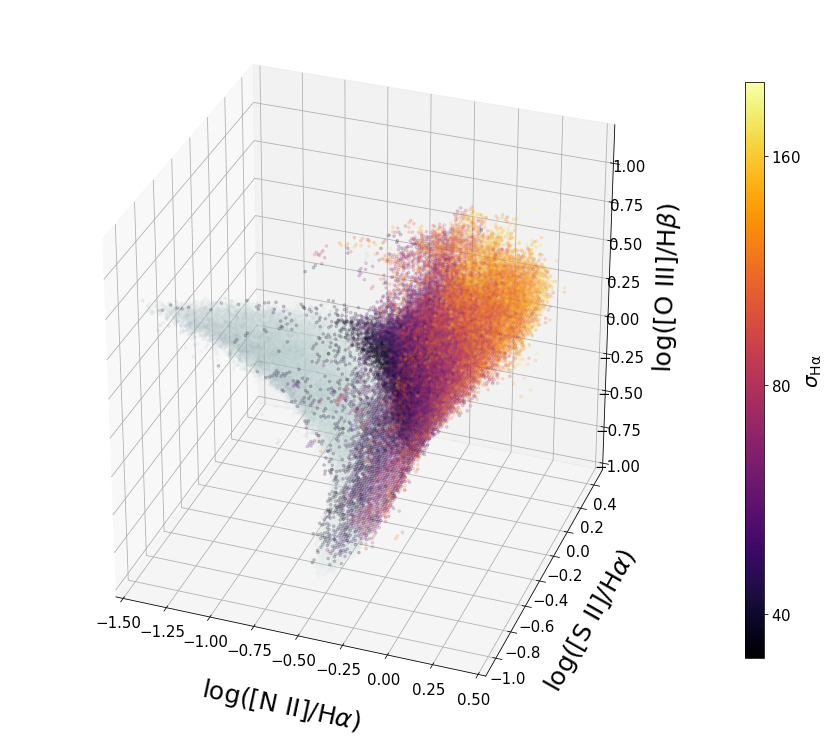}
    \includegraphics[width=0.33\textwidth]{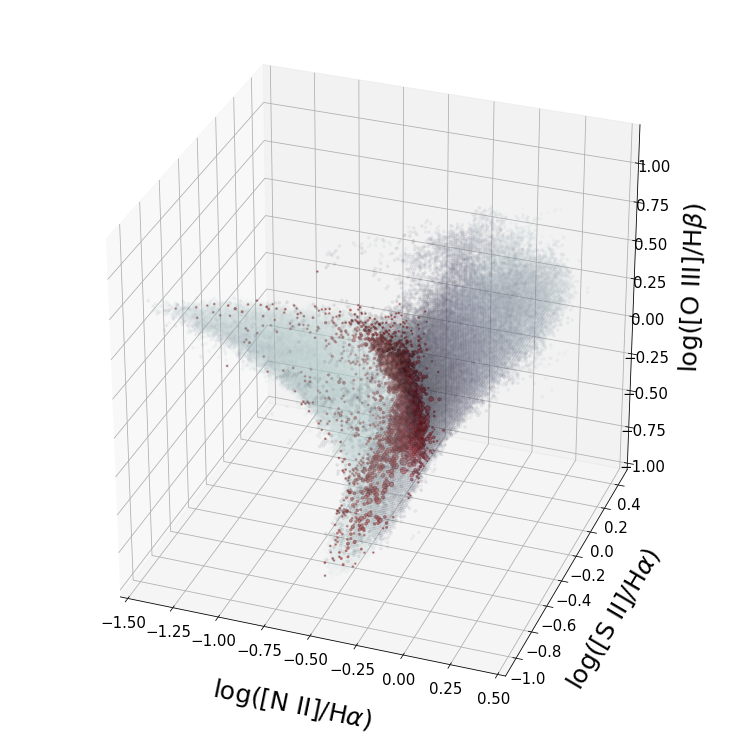}
    \caption{Left: 3D BPT diagram viewed at an elevation angle of $30^{\circ}$. Data are binned in the 3D line-ratio space, and the bins are colored coded according to their sigma-clipped mean $\sigma_{\Ha}$ values. While a total of $75\times75\times75$ bins are constructed, only bins made up of a minimum number of 5 spaxels are shown. An animated element shows the diagram rotating through different viewing angles.  Middle: bins with $\sigma_{\Ha}<35$ \kms\ are made transparent and only bins with $\sigma_{\Ha}>35$ \kms\ are colored coded. Right: bins with 
    $\sigma_{\Ha} = 33$ -- 37 \kms\ are shown in red, of which the displayed sizes are set to be proportional to the number of spaxels inside the bins. The rest of the bins are made transparent.}
    \label{fig:3d_BPT}
\end{figure*}

%Thus far we have restricted our analysis to two-dimensional diagnostic diagrams, but as discussed by multiple authors \citep[e.g.,][]{vogt14, ji20}, photoionization models actually define a hypersurface in the multidimensional line-ratio space, and many of the artifacts that we observe in the \redtxt{standard BPT diagrams} can be understood in terms of the projection of this surface onto a given plane \citep[e.g. Fig.~11 of the review by][]{kewley19}. 

In Figure~\ref{fig:3d_BPT}, we plot the distribution of our data in a 3D line-ratio space spanned by N2, S2, and R3. The data points are binned in this 3D space, and the color coding shows the sigma-clipped mean $\sigma_{\Ha}$ of the spaxels inside each bin. The left panel shows a clear color gradient in the data distribution, where $\sigma_{\Ha}$ increases from the SF locus to the AGN region.
%with a dividing $\sigma_{\Ha}$ close to 35 \kms. Regions with $\sigma_{\Ha} > \sigma_{\rm lim}=35$ \kms\ are clearly separated from the regions with $\sigma_{\Ha} < \sigma_{\rm lim}$, as can be seen in the middle panel. 
We note that the high $\sigma_{\Ha}$ bins continuously extend to low S2 values. When projected to two dimensions (e.g., Figure \ref{bpt1.fig}), the high $\sigma_{\Ha}$ spaxels
that overlap with the main star-forming cold-dispersion sequence are not numerous
enough to significantly affect the average.  At the low S2 end of the distribution
however there are substantially fewer cold star-forming spaxels,
and the high $\sigma_{\Ha}$ spaxels therefore give
rise to what appears to be an isolated `S2 bump' in the average velocity
dispersion and stellar population properties around
S2 $ = -0.8$ and R3 $ = 0$.
From a 3D point of view, at high metallicity the \ntwo/\Ha\ BPT diagram is thus a cleaner projection that better conserves the gradient of gas velocity dispersion. This arises because the photoionization model surface that describes the SF region is more edge-on in the \ntwo/\Ha\ diagram at high metallicity, and more face-on in the \stwo/\Ha\ diagram. As a result, the composite regions with high metallicities are able to extend further below the demarcation line in the \stwo/\Ha\ diagram. Although it is less obvious in the \ntwo/\Ha\ diagram, we also see spaxels of relatively high $\sigma_{\Ha}$ below the demarcation line. Therefore, we want to find a kinematically defined dividing surface in 3D that better isolates the dynamically cold component. With such a surface, we can further construct a 2D projection that minimizes the projected area of the surface.  This projection will help us visualize the separation of different kinematic components.
% DRL: Removing reference to figure, which I agree is confusing
%(e.g., the left and middle panels of Figure~\ref{bpt_photomodels.fig}). 
%However, the situation is different at low metallicity, for which neither of these 2D projections present the separation clearly and a better viewing angle is required for this intrinsically high dimensional space.

In \S \ref{defining.sec}, we used groups of spaxels offset in line ratio
space from the \citet{k03} relation to define a dynamically cold sample of spaxels,
and found that the $1\sigma$ upper boundary on this sample lay at about 35 \kms.
We now perform a similar calculation in 3D, computing the velocity dispersion
as a function of the 3D distance from the
star-forming photoionization model reference surface presented by \citet{ji20}.\footnote{As before with the \citet{k03} relation, our results do not
depend strongly on the choice of the model grid so long as it provides a 
reasonable match to the overall shape of the star-forming region.}
Once again, we find that the $1\sigma$ wing of the cold LOSVD is roughly $35~\kms$; 
the middle and right panels of Figure~\ref{fig:3d_BPT} show the locations of the 3D bins with $\sigma_{\Ha} \sim~35~\kms$ and the regions they separate in 3D.

%To construct such a dividing surface in 3D, we look for bins having average $\sigma_{\Ha}$ close to $35~\kms$ (i.e., similar to the $1\sigma$ criteria that we
%initially used to define our 2D relations in \S \ref{defining.sec}).
%The reason we choose this value is that back in 2D, we find the dynamically cold components in all standard BPT diagrams tend to have a $\sigma_{\Ha}$ distribution peaking around $24~\kms$ with a $1\sigma$ extreme of $35~\kms$. Although the exact value of this $1\sigma$ wing might well depend on the statistical method we use to compute the bin average, it should be close to our sigma-clipped result of $35~\kms$ within the uncertainty. 
%To ensure that this definition is still reasonable in 3D, we perform a similar calculation by looking at the histograms of $\sigma_{\Ha}$ for individual spaxels binned according to their distance to a reference surface in 3D. The reference surface is chosen to be a SF photoionization model surface with varying metallicity and ionization parameter \citep[see][for the details of the model]{ji20}. But we note that the result does not depend on the choice of model grid as long as it provides a reasonable match to the overall shape of the SF region in 3D. We find that the $1\sigma$ wing of the cold LOSVD is again roughly $35~\kms$, which ensures that we can keep using this definition for the 3D dividing surface. The middle and right panels of Figure~\ref{fig:3d_BPT} show the locations of the 3D bins with $\sigma_{\Ha} \sim~35~\kms$ and the regions they separate in 3D.}

\begin{figure}
\epsscale{1.1}
\plotone{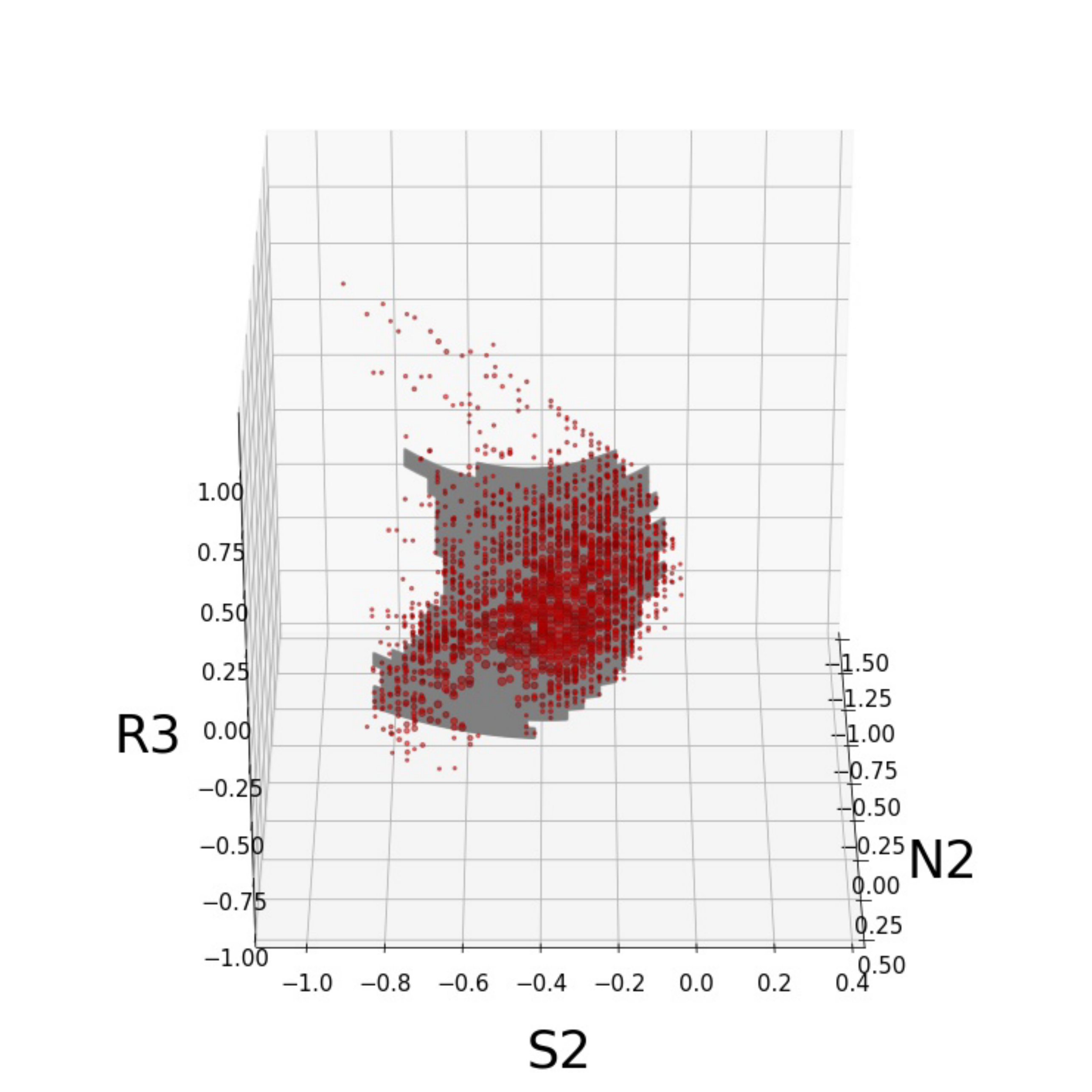}
\caption{Polynomial surface (grey) that fits the 3D bins (red) used to define the dividing surface for cold and warm kinematic components. The polynomial surface is cut so that only the part that covers the middle 99\% of the data along all three axes remains. The displayed sizes of the bins are scaled according to the number of spaxels they contain.
}
\label{polysurf.fig}
\end{figure}

By analogy to our approach in 2D, we then use an analytical function of
the form
\begin{align}
    N2 &= \Sigma ^2_{i=0} \Sigma ^2_{j=0} C_{ij} S2^i R3^j,
\end{align}
to trace the 35 \kms\ iso-dispersion surface in 3D.  In detail,
we use a fast least trimmed squares algorithm to fit 
3D bins with $33~\kms <\sigma_{\Ha} <37~\kms$
so as to minimize the influence of outliers \citep{rousseeuw06, cappellari13}.
To better reflect the distribution of the observed data in 3D, we only fit the part of the surface that continuously covers the middle 99\% of the 3D bins along all three axes (we additionally ensure that the surface behaves well when it is extrapolated outside the region covered by the MaNGA data). 
The final coefficients of this polynomial surface are given in Table~\ref{coef.table},
and illustrated in Figure~\ref{polysurf.fig}.

\begin{table}
    \centering
    \caption{Coefficients of the polynomial surface}
    \label{coef.table}
    \begin{tabular}{ |c|c|c|c| }
        \hline
        $C_{ij}$ & $j=0$ & 1 & 2 \\
        \hline
        $i=0$ & $-0.7362$ & $-0.6464$ & $-0.3036$ \\
        \hline
        1 & $-1.7567$ & $-1.2338$ & $0.4533$ \\ 
        \hline
        2 & $-2.0170$ & $-1.3520$ & $0.3177$ \\ 
 %       $j=0$ & $-0.7362$ & $-1.7567$ & $-2.0170$ \\
 %       \hline
 %       1 & $-0.6464$ & $-1.2338$ & $-1.3520$ \\ 
 %       \hline
 %       2 & $-0.3036$ & $0.4533$ & $0.3177$ \\ 
        \hline
    \end{tabular}
\end{table}

\begin{figure*}
\plotone{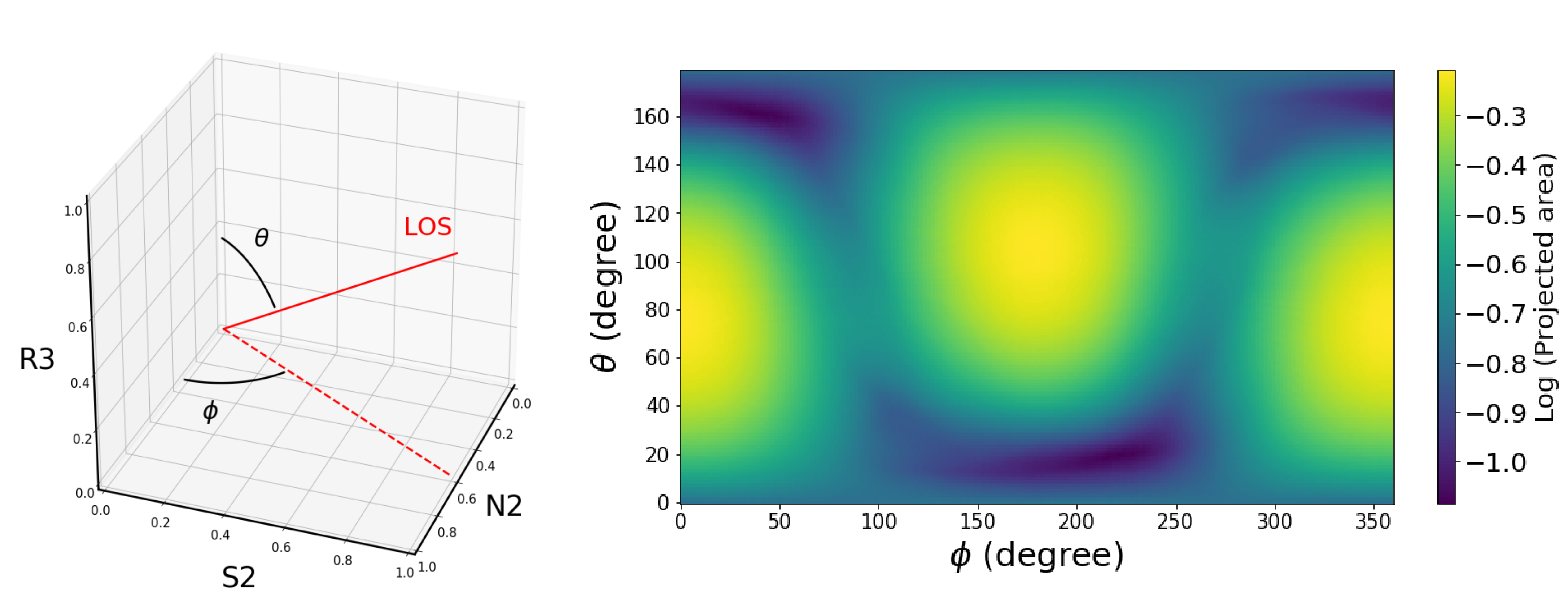}
%    \centering
%    \includegraphics[width=0.33\textwidth]{3d_bpt/ro_angles.png}
%    \includegraphics[width=0.6\textwidth]{3d_bpt/min_polysurf3337ltsc.png}
    \caption{Left panel: definition of the viewing angles in the 3D space. Right panel: logarithm of the projected area of the polynomial dividing surface as a function of the two viewing angles in 3D. The two degenerate global minima occur at $(\theta , \phi)=(19^{\circ}, 215^{\circ})$ and $(\theta , \phi)=(161^{\circ}, 35^{\circ})$.
}
    \label{porjarea.fig}
\end{figure*}

\begin{figure}
\epsscale{1.1}
\plotone{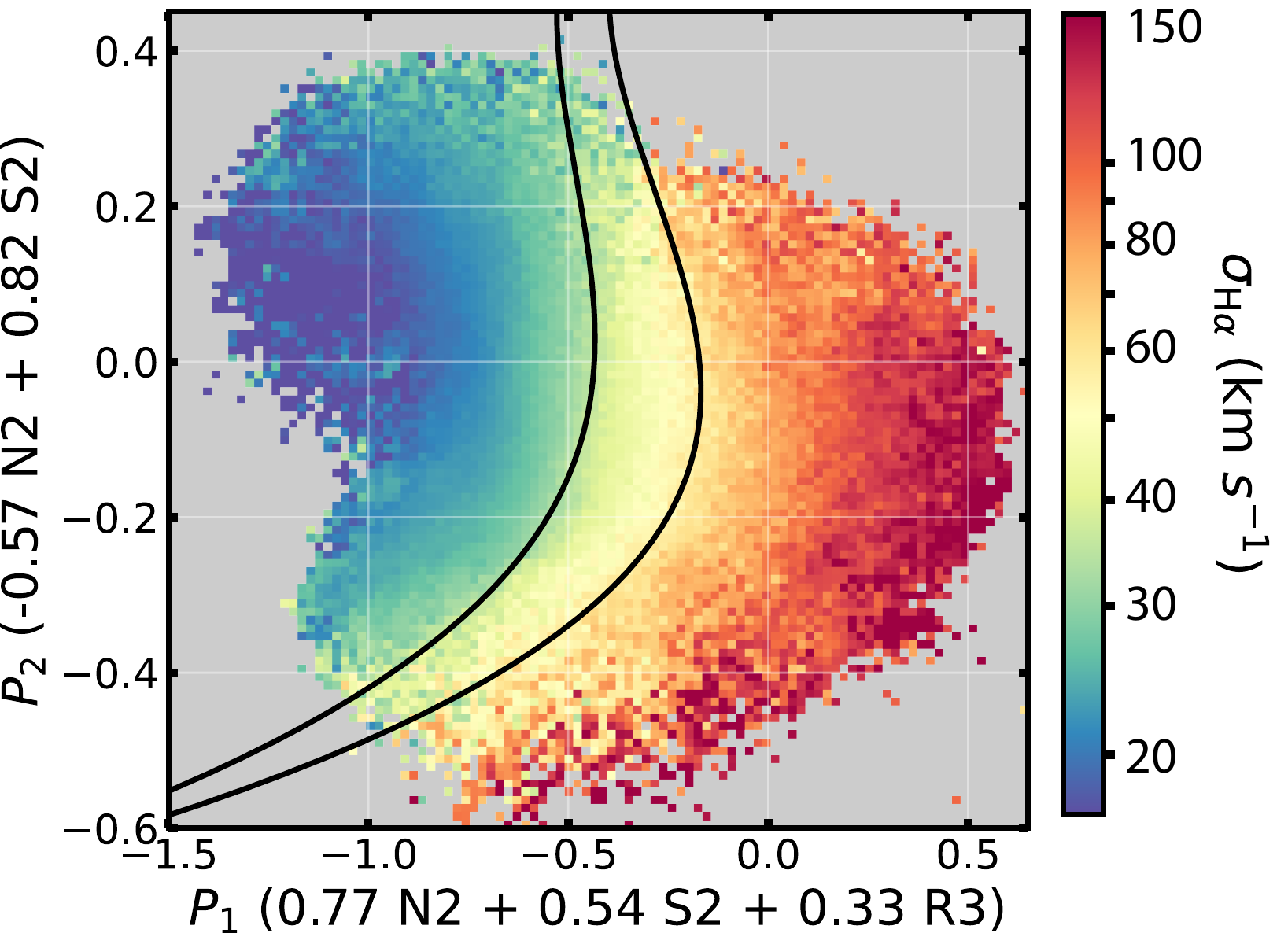}
\caption{Sigma-clipped mean \Ha\ velocity dispersion of the MaNGA spaxels
in our kinematically-defined $P_1 - P_2$ line ratio space that makes the 3D
iso-dispersion surface closest to edge-on.}
\label{kin_proj.fig}
\end{figure}

%To better demonstrate the advantage of using this dividing surface, we derive a 2D projection that helps to visualize the different kinematic components we see in 3D. 
Following \cite{ji20}, we next look for 
the viewing angles that make this 3D surface appear closest to edge-on
by minimizing its projected 2D area along different lines of sight.
Each line of sight is described by the polar angle $\theta$ and the azimuthal angle $\phi$ (in the right-handed coordinate system N2-S2-R3
shown in Figure \ref{porjarea.fig}, so that $\theta$ is defined relative to the positive direction of R3, and $\phi$ is defined relative to the positive direction of N2). We vary $\theta$ from $0^{\circ}$ to $180^{\circ}$ and $\phi$ from $0^{\circ}$ to $360^{\circ}$ in increments of $1^{\circ}$. The results are shown in Figure~\ref{porjarea.fig}, and indicate a clear valley of minima indicating the rotational symmetry of the polynomial surface. The global minima lie at $(\theta , \phi)=(19^{\circ}, 215^{\circ})$ and the rotationally degenerate point $(\theta , \phi)=(161^{\circ}, 35^{\circ})$ which specifies the same viewing axis
with the opposite sign. Figure~\ref{kin_proj.fig} shows the 2D projection corresponding to $(\theta , \phi)=(19^{\circ}, 215^{\circ})$. Since we have rotational freedom about the viewing axis, we choose the vertical axis of the projection, $P_{2}$ to be linear combinations of N2 and S2, while the horizontal axis, $P_{1}$ is a linear combination of N2, S2, and R3. These projections are given by 
\begin{align}
    P_1 &= 0.77 N2 + 0.54 S2 + 0.33 R3,
\end{align}
and
\begin{align}
    P_2 &= -0.57 N2 + 0.82 S2.
\end{align}

\begin{figure*}
    \centering
    \includegraphics[width=.95\textwidth]{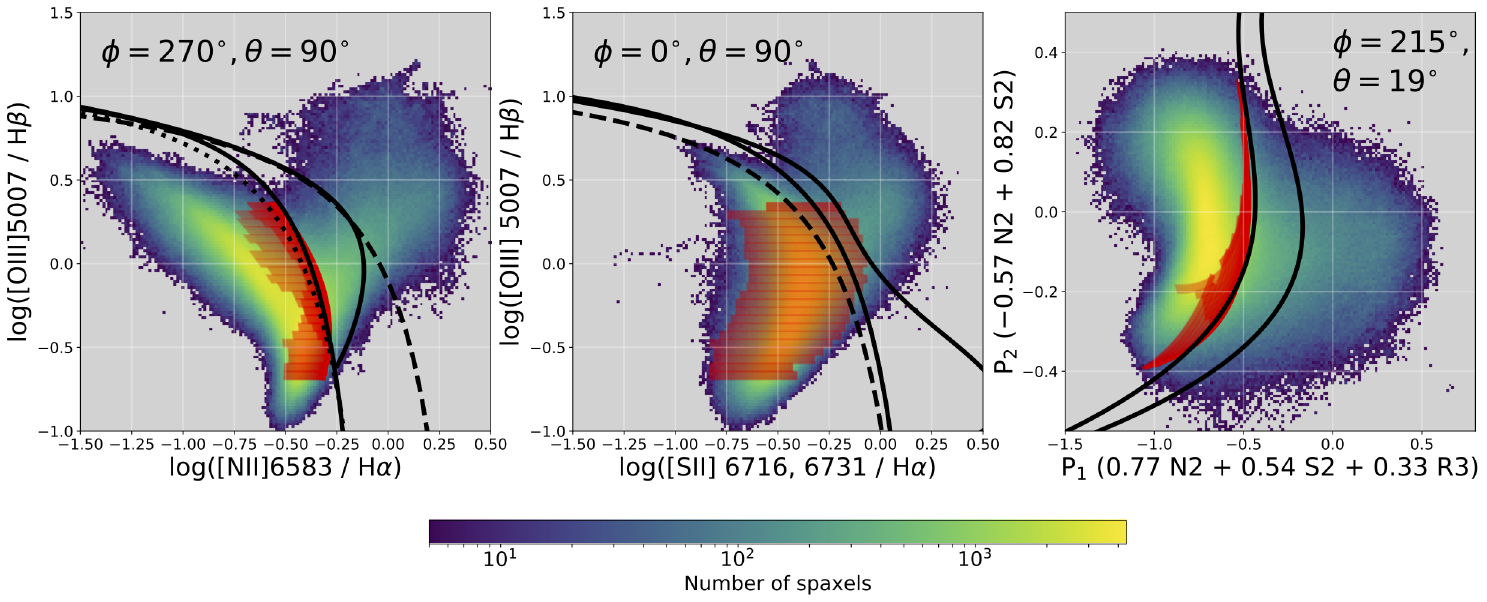}
    \caption{Density distributions of MaNGA spaxels in different projections of the three-dimensional line-ratio space. The polar coordinates $(\phi, \theta)$ are used to define the viewing angle, where $\phi$ is the angle with respect to the positive direction of the log(\ntwo/\Ha) axis, and $\theta$ is the angle with respect to the positive direction of the log(\othree/\Hb) axis. Red points represent the projected polynomial dividing surface defined in 3D.  Solid black lines represent
    our kinematic $1\sigma$ and $3\sigma$ demarcation lines, dashed lines are from
    \citet{kewley01}, and dotted lines are from \citet{k03}.
    %Red points are projected 3D bins with $33~{\rm km/s}<\sigma_{\Ha}<37~{\rm km/s}$. The size of each point is proportional to the number of spaxels inside the bin. Left panel: the N2 diagram with the demarcation lines defined in \cite{k03} (the dotted line), \cite{kewley01} (the dashed line), and this work (solid lines). Middle panel: the S2 diagram with the demarcation lines defined in \cite{kewley01} (the dashed line). The kinematic demarcation is plotted as the solid lines. Right panel: the empirically derived projection that minimized the projected kinematic dividing surface based on visual inspection. The two axes, $P_{1,k}$ and $P_2 = P_{2,k}$ are linear combinations of log(\ntwo/\Ha), log(\stwo/\Ha), and log(\othree/\Hb) (which are denoted as N2, S2, and R3 respectively). The dotted-dashed line represents a fit to the projected surface in the 3D space that corresponds to a maximum AGN contamination of 10\% for the SF regions, as defined in \cite{ji20}. The solid lines are the kinematic demarcations derived as in \S \ref{defining.sec}. 
    }
    \label{3proj.fig}
\end{figure*}

As illustrated in Figure \ref{3proj.fig} the 3D polynomial iso-dispersion
surface is almost face-on in the S2-R3 diagram, corresponding to 
a large fraction of spaxels with high $\sigma_{\Ha}$ (i.e., the `S2 bump') 
that are mixed with the dynamically cold component. 
%The edge of the surface roughly reaches our $3\sigma$ demarcation line defined in \S \ref{defining.sec}. 
In the N2-R3 diagram, the outer envelope of the 3D surface is well reproduced by
our $1\sigma$ demarcation line. Nonetheless, the surface still partly extends to the
left of this demarcation line, and as a result there will be some fraction of
spaxels misclassified as the cold-disk component in the N2-R3 diagram as well.
In contrast, the 3D surface has the narrowest projection of all in our
kinematically defined $P_1 - P_2$ frame, overlapping very little with the 
H\,{\sc ii} regions in the left-hand side of the figure.

In Figure~\ref{kin_proj.fig}, we plot the sigma-clipped mean $\sigma_{\Ha}$ of our data in this projection.
The mean velocity dispersion shows a regular well-defined gradient, and we follow \S \ref{defining.sec} in deriving corresponding
relations that bound the $1\sigma$ and 3$\sigma$ limits of the cold-disk
sequence as 
\begin{equation}
\label{projfit_1sig.eqn}
P_1 = 2.597 P_2^3 -1.865 P_2^2 +0.105 P_2 -0.435,
\end{equation}
and
\begin{equation}
\label{projfit_3sig.eqn}
P_1 = 3.4 P_2^3 -2.233 P_2^2 -0.184 P_2 -0.172.
\end{equation}
%The $1\sigma$ demarcation derived in this 2D projection is largely consistent with the directly projected 3D bins. 
As detailed in Table \ref{purity.table}, we find that the corresponding 
dynamically-cold purity of the spaxels selected in such a manner is 83\%,
with a completeness of 96\%.  The $P_1 - P_2$ selection method is therefore comparable to the N2-R3 method in terms of completeness, while both have
significantly better purity than the S2-R3 and O1-R3 methods.

We note that this `best' kinematic projection is not uniquely defined considering the uncertainties associated with the derivation, and other projections with line of sight near the local minima shown in Figure~\ref{porjarea.fig} might give equally good results. Similarly, it differs by about $20^{\circ}$ from the best projection angle derived
by \citet{ji20} using theoretical photionization models.
One explanation for this difference may be that \cite{ji20} defined their projection as the common normal plane of two representative patches on their SF-ionized model and AGN-ionized model, which effectively minimizes the projected areas of both of their model surfaces at the same time. The result might be different if only the SF-ionized model was considered.  Alternatively, if we had chosen a value other than $\sigma_{\Ha} = 35$ \kms\ for defining the iso-dispersion surface we would have plausibly derived
a different optimal projection.  Likewise, the 
photoionization models used by \citet{ji20} may need to be improved as the model-based
surface may have difficulty accurately describing very metal rich clouds.
Finally, the kinematic threshold separating regions dominated by different
ionization mechanisms may simply not be constant, and 
a fixed proportion of contamination from spatially close AGN-ionized regions or DIG could translate into slightly different velocity dispersions for composite regions in galaxies with different stellar masses. The connection between the theoretically-derived SF boundary and the empirically-derived one presented here
therefore cannot be easily understood without a consistent treatment of gas ionization properties and kinematics. Future work with sophisticated photoionization models based on hydrodynamical modeling of the gas structure might help to bridge the two methods.

\section{Summary}
\label{summary.sec}

We have presented the first
large-scale IFU analysis of the distribution of ionized gas velocity dispersions
in a representative sample of 9149 galaxies at median redshift $z = 0.04$.
As we demonstrated in \S \ref{defining.sec}, velocity dispersion correlates strongly
with the observed intensity ratios between strong nebular emission lines
such as \Ha, \othree, \ntwo, \stwo, and \oone.  By virtue of this correlation,
there exist well-defined regions in the multidimensional line ratio space
for which the ionized gas is either dominated by a dynamically cold
component with $\sigma_{\Ha} \approx 24 \pm 11$ \kms,
or a dynamically warm component with a much broader distribution of $\sigma_{\Ha}$
extending up to 200 \kms.
The dynamically cold sequence corresponds closely to the regions covered by traditional
stellar photoionization models, and we therefore identify it as representing star formation
in HII regions embedded in galactic thin disks.

Given this strong correlation between the
{\it sources} of the ionizing photons and the dynamical properties of the 
gas {\it illuminated} by these photons, we therefore used our observed isodispersion
contours to construct a series of equations (Eqns. \ref{1sig_n2.eqn} - \ref{3sig_o1.eqn})
defining the $1\sigma$ and $3\sigma$ boundaries between star-forming spaxels and those
dominated by AGN and/or LI(N)ER-like emission.
While these new boundaries are rooted in the observed ionized gas velocity dispersion, we
have demonstrated in \S \ref{properties.sec} that they correspond closely to 
well-studied trends in
other physical observables
as well.  The star-forming population has predominantly young ages and high star formation rates
implied by the low D$_n$4000 and high \Ha\ equivalent width respectively, and ionized gas velocity
dispersions that are 40\% or less of the corresponding stellar velocity dispersion.  In contrast,
the LI(N)ER sequence is much older and has lower \Ha\ equivalent width, with gas dispersions
$\sim 70$\% of the stellar velocity dispersion, consistent with expectations for a diffuse warm ionized medium.  The AGN sequence in turn is intermediate in population age and \Ha\ equivalent width, but
exhibits gas dispersions comparable to or in excess of the stellar velocity dispersion, indicative
of radiative shocks and non-equilibrium gas flows.
In practice the gas in each spaxel is by no means exclusively ionized by a single mechanism, and the Intermediate region between these sequences is therefore
likely a combination of a traditional mixing sequence augmented by the rise in contributions
from diffuse ionized gas in older stellar populations.

We explored the importance of selection effects
to our conclusions in \S \ref{selecteffect.sec}, the most crucial of which
is the galactocentric radius of the spaxel sample.
As we demonstrated in \S \ref{sdss1.sec}, there is a large population of low-metallicity
rapidly star-forming spaxels at distance of $1-2 R_{\rm eff}$ in the MaNGA sample that is
absent or severely underrepresented when we apply radial selection cuts designed to mimic
the fiber aperture coverage of the original SDSS-I spectroscopic survey.  These sample
differences are almost entirely responsible for the offsets between our new dynamically-defined 
sequences and the corresponding relations
defined by \citet{kewley01} and \citet{k03}.

We additionally provided theoretical support for our new classifications using
updated photoionization models in \S \ref{cloudy.sec}; these updated models are capable
of explaining the revised upper boundaries of both the \ntwo/\Ha\ and \stwo/\Ha\ dynamically cold sequences.  In contrast
the revised \oone/\Ha\ boundary cannot yet be explained theoretically, either because of
observational contamination
from diffuse ionized gas or inadequacies in theoretical models.

Finally, we presented a multidimensional view of the diagnostic line ratio diagrams, 
noting how the folding of photoionization models in this space naturally leads to both
an upper envelope of the star-forming sequence and artifacts when projected into
two dimensions.  For instance, the 3D surface defined by the N2, S2, and R3 
line ratios exhibits a tail to the LI(N)ER sequence that (when projected into the S2 vs
R3 space) manifests as a `bump' of old, high-$\sigma_{\Ha}$ spaxels below the traditional
star-forming sequence.  By allowing for a rotation of the projection angle of this 3D surface
we have determined a new two-dimensional projection that effectively minimizes such 
contamination (albeit comparably to the usual N2-R3 projection) 
and results in a strong and systematic gradient of $\sigma_{\Ha}$ 
across the projected line ratio space.

%Renbin:In the end, line ratios are still going to be more directly tied to 
%the ionization mechanism, perhaps what we are truly learning is how velocity dispersion varies 
%depending on different physical properties, with ionization mechanism being one of them. Ionization
%mechanism variation provides a varying flashlight that highlights different gas components in different
%cases. But the dispersion of each gas component could also change depending on the local gravity,  and 
%other turbulence-driving factors. Perhaps these other things could make the constant velocity dispersion 
%curve to be different from the true ionization-transition curve. 

\acknowledgments

MAB acknowledges NSF Awards AST-1517006 and AST-1814682.

Funding for the Sloan Digital Sky Survey IV has been provided by the Alfred P. Sloan Foundation, the U.S. Department of Energy Office of Science, and the Participating Institutions. SDSS-IV acknowledges
support and resources from the Center for High-Performance Computing at
the University of Utah. The SDSS web site is www.sdss.org.

SDSS-IV is managed by the Astrophysical Research Consortium for the 
Participating Institutions of the SDSS Collaboration including the 
Brazilian Participation Group, the Carnegie Institution for Science, 
Carnegie Mellon University, the Chilean Participation Group, the French Participation Group, Harvard-Smithsonian Center for Astrophysics, 
Instituto de Astrof\'isica de Canarias, The Johns Hopkins University, Kavli Institute for the Physics and Mathematics of the Universe (IPMU) / 
University of Tokyo, the Korean Participation Group, Lawrence Berkeley National Laboratory, 
Leibniz Institut f\"ur Astrophysik Potsdam (AIP),  
Max-Planck-Institut f\"ur Astronomie (MPIA Heidelberg), 
Max-Planck-Institut f\"ur Astrophysik (MPA Garching), 
Max-Planck-Institut f\"ur Extraterrestrische Physik (MPE), 
National Astronomical Observatories of China, New Mexico State University, 
New York University, University of Notre Dame, 
Observat\'ario Nacional / MCTI, The Ohio State University, 
Pennsylvania State University, Shanghai Astronomical Observatory, 
United Kingdom Participation Group,
Universidad Nacional Aut\'onoma de M\'exico, University of Arizona, 
University of Colorado Boulder, University of Oxford, University of Portsmouth, 
University of Utah, University of Virginia, University of Washington, University of Wisconsin, 
Vanderbilt University, and Yale University.

%\appendix
%
%\section{DAP reliability}
%
%Plot stacked spectra in various regions of BPT space, along with DAP stacked model fits to demonstrate
%overall reliability.

\end{document}